\documentclass[preprint, 3p, number]{elsarticle}

\usepackage{latexsym}
\usepackage[pdftex]{color}
\usepackage{amsthm}
\usepackage{graphicx}
\usepackage{amssymb}
\usepackage{subfig}
\usepackage{amsmath}
\usepackage{float}
\usepackage{parskip}
\usepackage{textcomp}
\usepackage[font=footnotesize,labelfont=bf]{caption}
\usepackage{array}
\usepackage{longtable}

\newtheorem{theorem}{Theorem}
\newproof{pf}{\bf{Proof}}
\numberwithin{equation}{section}
\numberwithin{figure}{section}
\newcommand{\sgn}{\mathop{\mathrm{sgn}}}
\newcommand{\BigO}[1]{\ensuremath{\operatorname{O}\bigl(#1\bigr)}}

\journal{}

\begin{document}
\begin{frontmatter}
\title{A nodal domain theorem for integrable billiards in two dimensions}
\author[rvt]{Rhine Samajdar}

\author[focal]{Sudhir R. Jain\corref{cor1}}
\ead{srjain@barc.gov.in}

\cortext[cor1]{Principal corresponding author \hspace{0.2cm} Phone: +91 22 25593589}
\address[rvt]{\textsl{Indian Institute of Science, Bangalore 560012, India.}}
\address[focal]{\textsl{Nuclear Physics Division, Bhabha Atomic Research Centre, Mumbai 400085, India.}}

\begin{abstract}
Eigenfunctions of integrable planar billiards are studied --- in particular, the number of nodal domains, $\nu$ of the eigenfunctions with Dirichlet boundary conditions are considered. The billiards for which the time-independent Schr\"{o}dinger equation (Helmholtz equation) is separable admit trivial expressions for the number of domains. Here, we discover that for all separable and non-separable integrable billiards,  $\nu $ satisfies certain difference equations. This has been possible because the eigenfunctions can be classified in families labelled by the same value of $m\mod kn$,  given a particular $k$, for a set of quantum numbers, $m, n$. Further, we observe that the patterns in a family are similar and the algebraic representation of the geometrical nodal patterns is found. Instances of this representation are explained in detail to understand the beauty of the patterns. This paper therefore presents a mathematical connection between integrable systems and difference equations.  

\end{abstract}


\end{frontmatter}

\setlength{\tabcolsep}{13.5pt}
\renewcommand{\arraystretch}{2.5}
\section{Introduction}
\noindent
The `particle in a box' has served as a model for understanding various phenomena in solid state and nuclear physics -- the theory of dynamical systems terms these as `billiards'. Studies of their energy spectra and eigenfunctions, and their connections with quantum chaos have been very fruitful and exciting. One of the properties of the eigenfunctions of these billiards is the organisation of regions with  positive and negative signs. These domains appear in rather complex forms \cite{courant}; their number displays a near-incomprehensible order if organised in increasing energy. We are familiar with domains that appear in a system which has two states or phases. For instance, in magnetic materials, there are regions of positively and negatively aligned spins. Their shapes and areas promise interesting statistical questions. There has been a lot of interest in studying the nodal domain statistics in recent times of billiards in two dimensions \cite{ss}. Here, we present a general result for the number of nodal domains of integrable plane polygonal billiards -- the geometries are rectangle, circle, ellipse, and triangles with angles (45, 45, 90), (30, 60, 90) and (60, 60, 60). The great interest in these systems emerges from the simplicity they seem to present, and their ubiquity in a large number of contexts. 

The Schr\"{o}dinger equation for a particle inside a rectangular box satisfying Dirichlet or Neumann boundary conditions can be immediately solved. The eigenfunctions are the well-known product of two sine or cosine functions and the nodal domains make a checkerboard. For the right isosceles and the equilateral triangle billiards, which are non-separable, the solutions are respectively two and three terms, each a product of two sinusoidal functions. For the equilateral triangle, we discovered a difference equation satisfied by the number of nodal domains, $\nu _{m, n}$ \cite{sj}, a discrete variable of the system. In this article, we show that $\nu _{m, n}$ for all the integrable billiards obey similar difference equations. This general result is amazing, in view of widely varied findings observed for these billiards about various statistical measures \cite{uzy2002,sankar,aronovitch}. Interestingly, recent investigations of integrable lattice systems \cite{levi} have also sought to probe the intricate connection between the theory of exactly integrable discrete systems and the formalism of difference equations.  

\section{General mathematical formulation}
Let ${\cal D} \subset \mathbb{R}^2$ be a compact, connected domain on a surface with a smooth Riemannian metric in two dimensions. Assuming Dirichlet conditions along the boundary $\partial {\cal D}$ and denoting the Laplace--Beltrami operator by $\nabla^2$, the eigenvalue problem is formulated as 
\begin{equation}
-\nabla^2  \psi_j = E_j  \psi_j \mbox{ and } \psi_j|_{\partial {\cal D}} = 0.
\end{equation}
A nodal domain of the wavefunction $\psi_j$ is a connected domain in ${\cal D}$ where $\psi_j \ne 0$, which therefore defines a maximally connected region wherein the function does not change sign. The subscript $j$ simply denotes the ordering of the spectrum such that $E_j \le E_{j+1}$. The importance of the nodal set arises from the fact that the sequence of the number of nodal domains of the eigenfunctions of the Schr\"{o}dinger equation not only bears significant geometric information about the system \cite{toth2} but also provides a new criterion for chaos in quantum mechanics \cite{uzy2002}. Hence, it may be reasonable to conjecture that the difference equations also encode the geometry of the system itself.

Any wavefunction $\psi_j$ on the domain ${\cal D}$ is characterised by two quantum numbers for manifolds on $\mathbb{R}^2$ which are represented hereafter as $m$ and $n$, unless specified otherwise, where $m, n \in \mathbb{N}$. Let $\nu_{m,n}$ denote the total number of nodal domains of the wavefunction $\psi_{m,n}$. Furthermore, let $R_{k, n}$ be an equivalence relation defined on the set of wavefunctions as 
\begin{equation}
R_{k, n} = \{ (\psi(m_1,n), \psi(m_2,n)) : m_1 \equiv m_2 (\mbox{mod } kn) \}.
\end{equation}
The relation $R_{k, n}$ defines a partition ${\cal P}$ of the set of wavefunctions into equivalence classes $[{\cal C}_{kn}]$ where ${\cal C}_{kn} = m \mbox{ mod } k\, n$. Here, the parameter $k$ represents the number of linearly independent terms of which the wavefunction $\psi_{m, n}$ is a sum. Consideration of the sequence of $\nu_{m,n}$ for wavefunctions that belong to the same class illustrates the rich structure of the difference equations that arise in two-dimensional integrable billiards. To this end, we introduce the forward difference operator, $\Delta_t {\cal F}(x_1, x_2)$, operating on the first index $x_1$ of a generic function ${\cal F}$, with a finite difference $t$, i.e. $\Delta_t {\cal F}(x_1, x_2) = {\cal F}(x_1+t, x_2) - {\cal F}(x_1, x_2)$.  \newline
\begin{theorem}
\textsl{If the metric space ${\cal D} \subset \mathbb{R}^2$ is integrable, then, in the absence of tiling, one of $\Delta_{kn} \nu\,(m,n) = \Phi(n) $ and $\Delta^2_{kn} \nu\,(m,n) = \Phi(n)\; \forall \, m, n$ holds for some $ \Phi: \mathbb{R} \rightarrow \mathbb{R}$, which is determined only by the geometry of the billiard.}
\label{one}
\end{theorem}
\begin{pf}
The statement can be easily demonstrated by verifying it individually for all possible integrable billiards on $\mathbb{R}^2$. The corresponding functions $\Phi(n)$ have been calculated in the respective sections.
\end{pf}
The major utility of this theorem lies in the fact that it presents a method to conveniently compute expressions for the total number of nodal domains, even for non-separable billiards, which may seem intractable otherwise. The reader is requested to note that the authors have taken the liberty of a slight abuse of notation in the sense that $\nu_{m, n}$ and $\nu (m, n)$ have both been used interchangeably to denote the number of nodal domains, depending on whether the quantity is regarded as a numeric variable or as a function, respectively.
\section{Separable billiards}
\noindent
Birkhoff conjectured that among all billiards inside smooth convex curves, ellipses are characterised by integrability of the billiard map \cite{birkhoff}.  Examples of systems that are both integrable and separable are presented by the rectangular, circular and elliptic billiards. Separability arises from the observation that the corresponding Schr\"{o}dinger equations on ${\cal D}$ are separable in certain coordinate systems \cite{morse}. Owing to the structure of the nodal pattern, which is formed by a grid of intersecting nodal lines for both the billiards, the expressions for $\nu_{m,n}$ are relatively uncomplicated being equal to the number of nodal crossings. The wavefunctions for both the billiards comprise of only a single term and hence $k = 1$, which implies that the equivalence class is defined as ${\cal C}_{1n} = m \mbox{ mod } n$.   Although such a construction of the class may seem artificial for the rectangular and circular billiards given the simple expressions for $\nu_{m,n}$, the formulation helps to naturally extend a seamless transition to the non-separable cases where no such exact formulae are known. In fact, the simplification obtained on approaching the nodal domain problem through the method of difference equations becomes perceptible for the elliptic billiard, for which the expression for $\nu_{m,n}$ is not trivially obvious.
\begin{figure} [H]
\begin{raggedleft}
\qquad \qquad
\subfloat[]{\scalebox{0.27}{\includegraphics{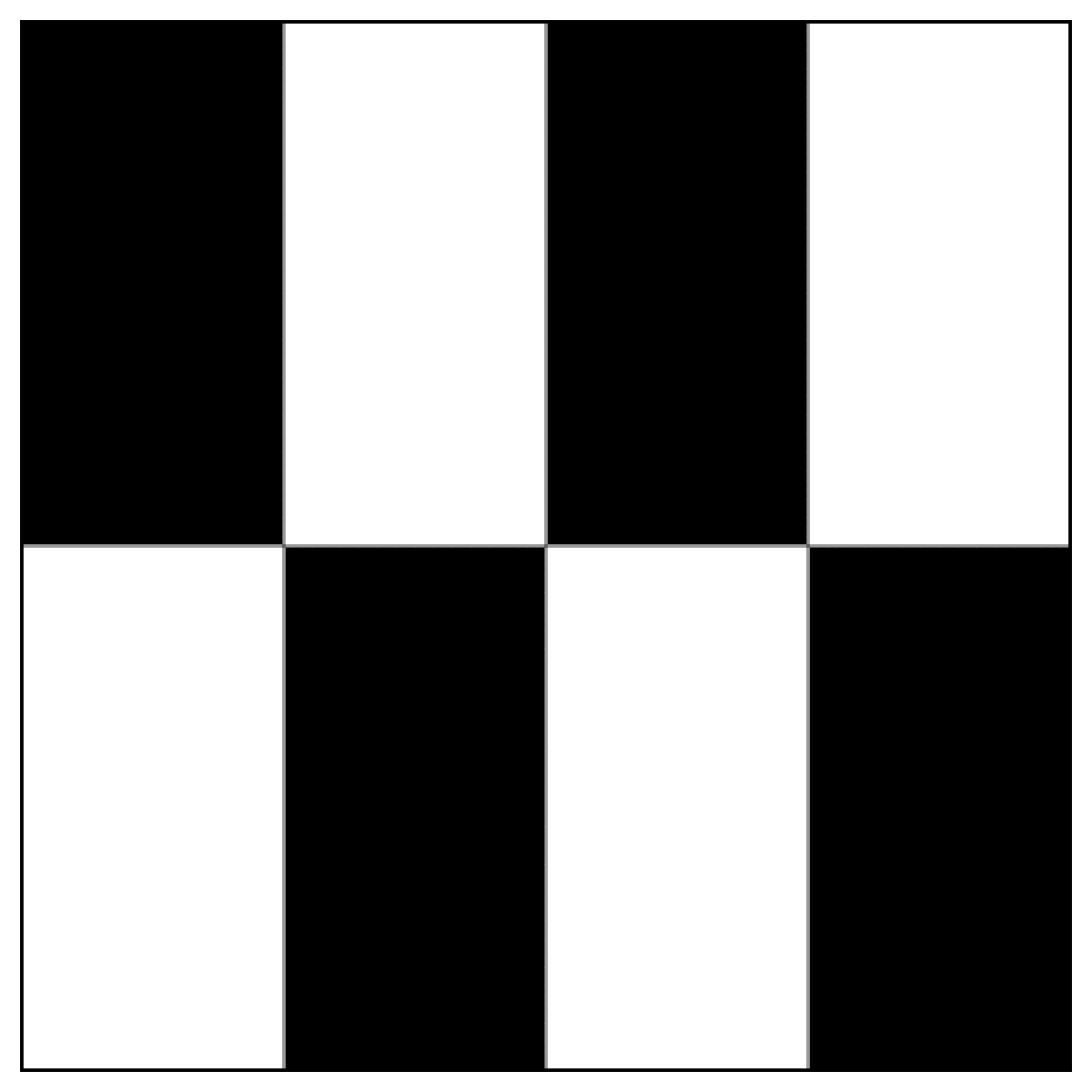}}}
\qquad\qquad \quad
\subfloat[]{\scalebox{0.27}{\includegraphics{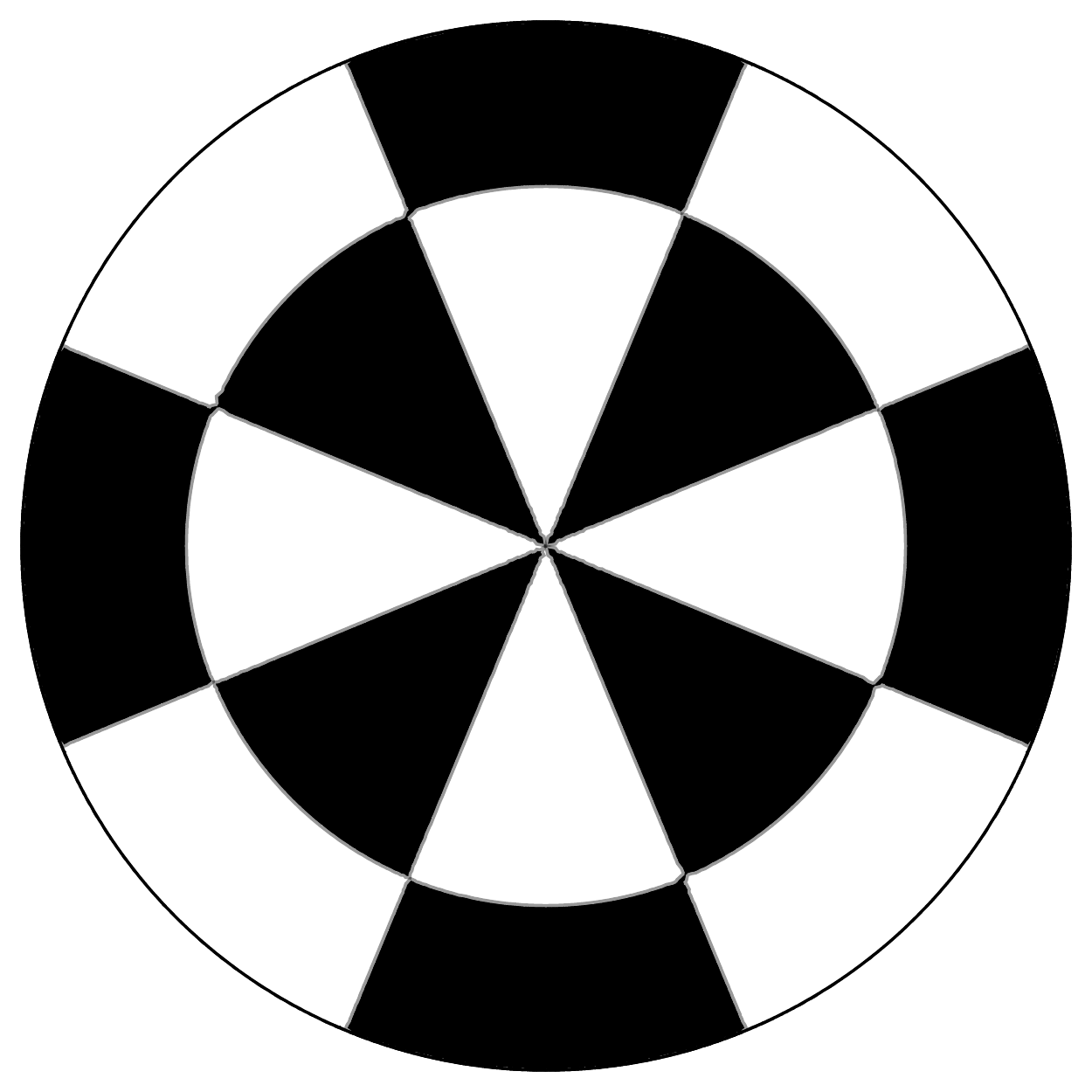}}}
\qquad\qquad\qquad
\subfloat[]{\includegraphics[scale = 0.370, trim = 7cm 10.3cm 7cm 15cm]{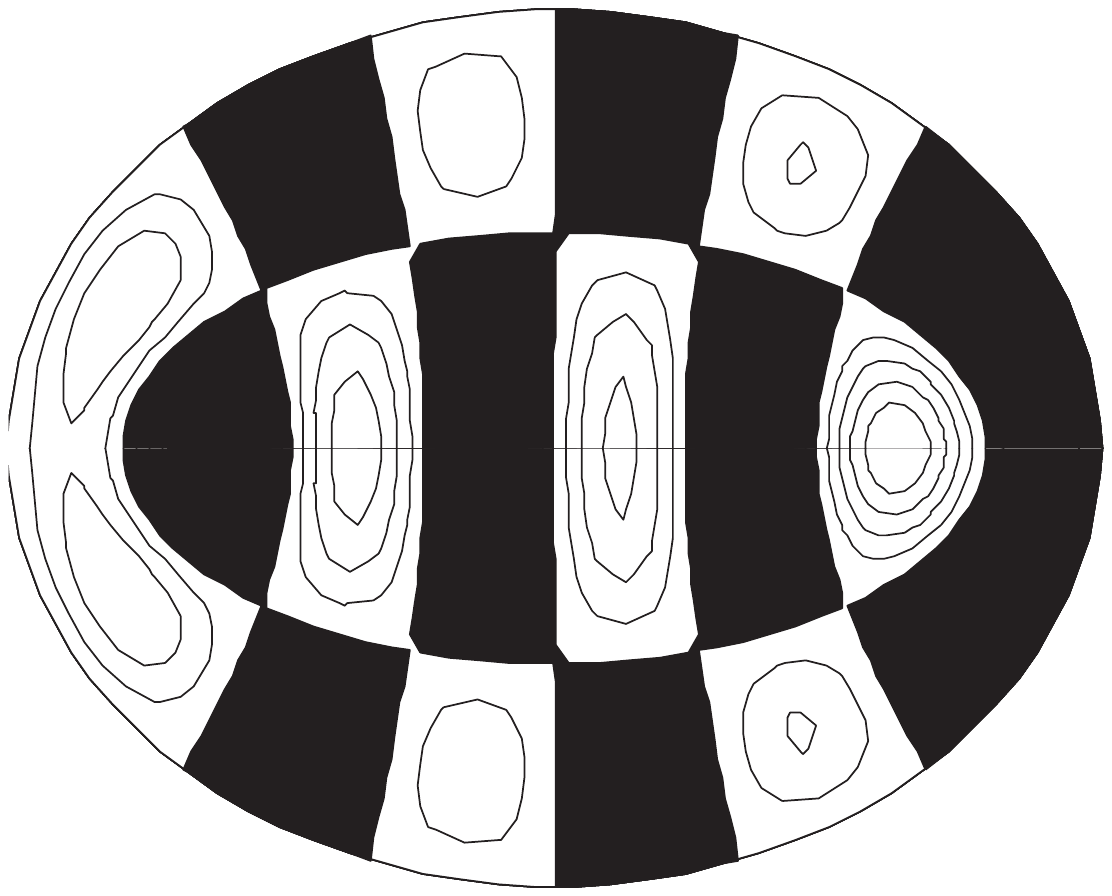}}
\end{raggedleft}
\caption{The `checkerboard' pattern of the nodal domains for (a) the rectangular billiard and (b) the circular billiard shown for the quantum numbers $m = 4$ and $n = 2$. The total number of nodal domains are $mn$ and $2mn$ for the rectangle and the circle respectively. The wavefunction is positive in the white areas and negative in the darkened regions. (c) The elliptic billiard (of eccentricity $\sqrt{2}$) displays a similar pattern of nodal domains as observed in the  + -- parity mode, symmetric about the X-axis (shown), plotted using the code developed by \cite{matlab}.}
\label{A1E}
\end{figure} 
In the following discussion, we only consider billiards with the imposition of Dirichlet boundary conditions, in which the wavefunctions vanish at the boundary, because such systems directly correspond to the model of a `particle in a box' through the physical assumption of an infinite potential outside the relevant region of the billiard. For Neumann boundary conditions, the generalisation of the results presented herein is trivial for separable integrable systems, since the same difference equations and expressions for $\nu_{m, n}$ hold. However, the extension of these relations to the non-separable integrable billiards when Neumann boundary conditions are imposed would certainly be an interesting exercise.
\subsection{The rectangle}
\noindent
The nodal domain distribution for rectangular billiards has been considered in detail in \cite{sankar}. Let ${\cal D} = \big[0,L_x\big] \times [0,L_y]$ be a rectangular drum such that $\alpha = L_x / L_y$. The eigenfunctions for the system are given by
\begin{equation}
\psi_{m, n} (x, y) = \sqrt{\frac{4}{L_x L_y}} \sin \bigg(\frac{m \pi x}{L_x} \bigg) \sin \bigg(\frac{n \pi y}{L_y} \bigg).
\end{equation}
Choosing $L_x = \pi$, the Dirichlet spectrum simplifies to $E = m^2 + \alpha^2 n^2$. Examination of the `checkerboard' pattern of nodal domains easily yields the number of nodal domains of a particular eigenfunction to be $\nu_{m,n} = mn$. The difference equation satisfied by this system is  
\begin{equation}
\Delta_{n} \nu\,(m,n) = \nu_{m+n, n} - \nu_{m, n} = n^2.
\label{rectangle}
\end{equation} 
The recurrence relation of Eq. \eqref{rectangle} can be solved analytically to obtain $\nu_{m, n} = mn + C_r$, where $C_r$ is a constant. Comparison with numerical results shows that $C_r$ = 0.
\subsection {The circle}
\noindent
The circle corresponds to a special case of an ellipse, which is an integrable billiard \cite{ellipse}. A circular domain of radius $R$ in two dimensions, physically corresponding to a circular infinite well, may be specified as ${\cal D} = \big\{(x, y): x^2 + y^2 \le R^2 \big \}$. In polar coordinates, the solutions of the Schr\"{o}dinger equation (for a particle of mass $\mu$) are separable into angular and radial components and are given by \cite{robinett1996, robinett} as
\begin{equation}
\psi_{m, n} (x, y)  = \frac {J_{m}(kr)}{\sqrt{\int_{0}^{R}\big[J_{m}(kr) \big]^2 r\, \mathrm dr }}.\frac{\cos m \theta}{\sqrt{\pi}};\:\: k = \sqrt{\frac{2\mu E}{\hbar^2}},
\label{eq:circ}
\end{equation}
where $J_m (z)$ denotes the cylindrical Bessel function of the first kind, which has non-divergent solutions as $z \rightarrow 0$. The energy spectrum for the system is therefore $E = \big[ z_{m, n}\big]^2 $, with $z_{m, n}$ representing the $n^{th}$ zero of the regular Bessel function $J_m(z)$ \cite{robinett}. Since the angular component of the eigenfunction, $\Theta_m (\theta) = e^{\mathrm{i} m \theta}$, is complex, the nodal domains are visualised for Eq. \eqref{eq:circ} by considering solely the real part of the wavefunction. The results derived in this section remain invariant on consideration of the imaginary part. Study of the nodal pattern, which bears structural similarity to that of the rectangle, indicates the existence of a comparable difference equation for the nodal counts as
\begin{equation}
\Delta_{n} \nu\,(m,n) = \nu_{m+n, n} - \nu_{m, n} = 2n^2 \hspace{1 cm} \mbox {if } m\ne0 .
\label{circle}
\end{equation} 
The analytical solution to this recurrence relation is nearly identical to that for Eq. \eqref{rectangle} except for a prefactor and is given by $\nu_{m, n} = 2mn + C_c$, where $C_c$ is a constant. On examining the numerically obtained number of nodal domains, it is observed that $C_c = 0$ and $\nu_{m, n}$ simplifies to $2mn$. When $m = 0$, the number of nodal domains is simply $n$ and the appropriate modification to Eq. \eqref{circle} would therefore read as $\Delta_{n} \nu\,(0,n) = 2n^2 - n$. 
{\small
\begin{center}
\begin{longtable}{c c c | c c | c c}
\hline
$m$ & $n$ & ${\cal C}_{1n} = m\mod n$ & $(\nu_{m,n})_{\textsl{Rectangle}}$ &  $\Delta_n \: \nu\,(m, n)$ & $(\nu_{m,n})_{\textsl{Circle}}$ & $\Delta_n \: \nu\,(m, n)$ \endhead \hline
14 & 11 & 3 & 154 &   --  & 308 &   --    \\
25 & 11 & 3 & 275 & 121 & 550 & 242 \\
36 & 11 & 3 & 396 & 121 & 792 & 242 \\
57 & 11 & 3 & 517 & 121 & 1034 & 242 \\
68 & 11 & 3 & 638 & 121 & 1276 & 242 \\
\hline
\caption{A tabular demonstration of the difference equations satisfied by the total number of nodal domains for the wavefunctions in a rectangle and a circle when the eigenfunctions belonging to the same class, defined by $m\mod n$, are arranged in increasing order of the quantum number $m$. It is easily observed that the rectangular and circular billiards differ in both $\nu_{m, n}$ and $\Delta_n \nu\,(m, n)$ only by a factor of 2.}
\end{longtable}
\end{center}
}
It is interesting to note that the Bessel function $J_m (z)$ tends to the standard trigonometric functions as $z \rightarrow \infty$, i.e. the function of the first kind resembles a sine or cosine curve, with a period that slowly shortens, eventually becoming $2\pi$. In the limit of $z \rightarrow \infty$, the boundary of a circle with a sufficiently large radius of curvature $\gg 1/k$ can be locally approximated by a straight line and thus the eigenfunctions, and consequently the nodal patterns, of the rectangular and circular billiards are intrinsically linked. The fact that the difference equations \eqref{rectangle} and \eqref{circle} retain this connection corroborates the non-triviality of these recurrence relations and seems to suggest the presence of a more fundamental underlying reason for their existence.

\subsection{The ellipse}
The notions introduced in the treatment of the circle are generalised for the elliptic billiard, which is separable in elliptic coordinates, $\xi$ and $\eta$, and the solutions thereof are described in terms of the radial and angular Mathieu functions \cite{traiber}. The radial and angular quantum numbers, $r$ and $\ell$ respectively, describing the system are defined in accordance with the notational convention for parity adopted in \cite{ellipse} as this is in agreement with the Einstein-Brillouin-Keller (EBK) quantisation for the symmetry reduced system. For the purpose of demonstration of the difference equations, we consider the states with even parity $++$ for which $\pi_x = 1 = \pi_y$ (refer to Figure 6 of \cite{ellipse}). For $\psi (r, \ell)_{\pi_x, \pi_y}$, we have
\begin{equation}
 \Delta_\ell \: \nu\,(r, \ell) = \nu_{r+\ell,\, \ell} - \nu_{r,\, \ell} = 4\ell^2,
\label{ellipse}
\end{equation} 
the solution to which is $\nu_{r, \ell} = \ell(4r+2) + 1$. Additionally, for the states where $\ell = 0$, it is to be noted that $\nu_{r,\, \ell} = r+1$, which trivially implies that  $\Delta_\ell \, \nu\,(r,\, 0) = 4\ell^2 = 0$. It is indeed straightforward to verify that this construction can be extended to the other parity combinations, $+ -$ (Figure \ref{A1E}.c), $- +$ and $ -- $. An interesting discussion on the hardening and softening of the pseudoradial and pseudoangular quanta respectively during the transition from the circle to the rectangle is presented in \cite{traiber} which underlines the inherent connection amongst these separable billiards and hence, between their `checkerboard' nodal patterns. 

\section{Non-separable billiards}
The eigenfunctions of the separable systems discussed in the previous section present a behaviour which is different from `generic' wavefunctions, the nodal lines of which, according to Uhlenbeck's hypothesis, do not intersect \cite{uhlenbeck}. The eigenfunctions of classically integrable but non-separable systems show avoided intersections and a few crossings of nodal lines \cite{monastra}, which greatly complicates the expression for $\nu_{m,n}$.

The primary focus of our discussions in the following sections with respect to the non-separable integrable billiards has exclusively been the non-tiling wavefunctions thereof. This apparent bias may be justified on the grounds that the number of nodal domains of a tiling state can be obtained by simply decomposing it into its constituent tiles and summing over the corresponding number in each individual non-tiling state. For instance, the phenomenon of tiling is most easily observable when $\gcd (m, n) = d \ne 1$ -- under such circumstances, $\nu_{m, n} = d^2 \nu_{m/d, n/d}$, since the nodal pattern of the wavefunction $\psi_{m, n}$ is composed of exactly $d^2$ tiles. Nevertheless, the validity of Theorem \ref{one} is retained under tiling circumstances. To illustrate this point, we consider, as previously, the situation when  $\gcd (m, n) = d \ne 1$, which implies that the state $(m/d, n/d)$ is irreducible into further tiles. Then, by the premise of Theorem \ref{one}, $\exists \, \Phi: \mathbb{R} \rightarrow \mathbb{R}$ such that either $\Delta_{kn} \nu\,(m,n) = \Phi(n) $ or $\Delta^2_{kn} \nu\,(m,n) = \Phi(n)$ is satisfied $\forall \, m, n$. It is easy to see that Theorem \ref{one} is satisfied, albeit with a different function $\hat{\Phi}= d^2 \Phi$ for the tiling states, since at least one of the following holds:
\begin{eqnarray*}
\\ \nu_{m+kn, n} - \nu_{m, n} &=& d^2 \big( \nu_{(m+kn)/d, n/d} - \nu_{m/d, n/d} \big) = d^2 \Phi, \\
\big(\nu_{m+2kn, n} - \nu_{m + kn, n} \big) - \big(\nu_{m+kn, n} - \nu_{m, n} \big)  &=& d^2\big(\nu_{(m+2kn)/d, n/d} - 2\nu_{(m + kn)/d, n/d} + \nu_{(m/d, n/d)} \big) = d^2 \Phi,
\end{eqnarray*} 
where the equality follows trivially from the fact that $m + kn$ is divisible by $d$.

\subsection{The right-angled isosceles triangle}
\noindent
\begin{figure} [H]\scriptsize
\begin{center}
\subfloat[]{\scalebox{0.3}{\includegraphics{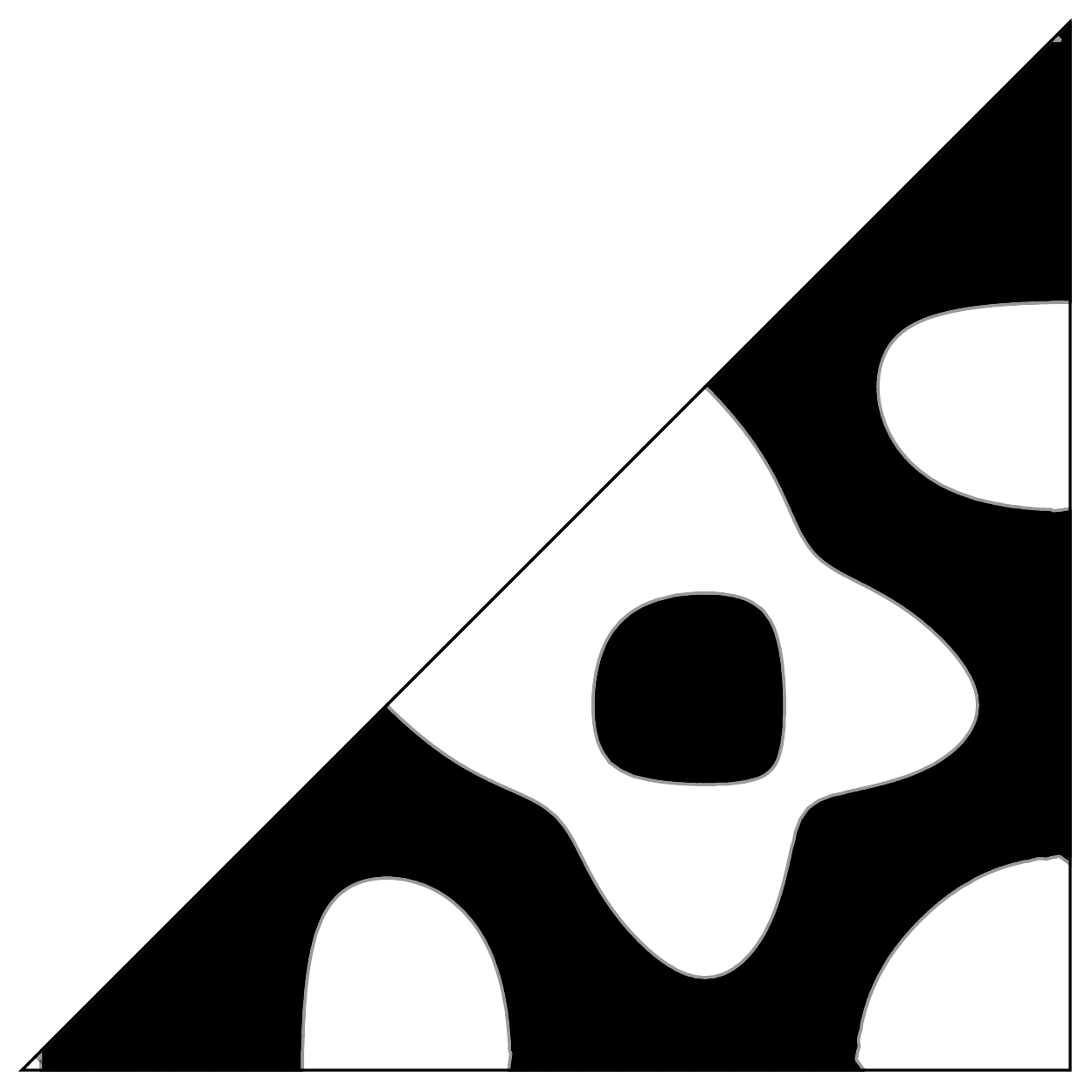}}}
\qquad \qquad
\subfloat[]{\scalebox{0.3}{\includegraphics{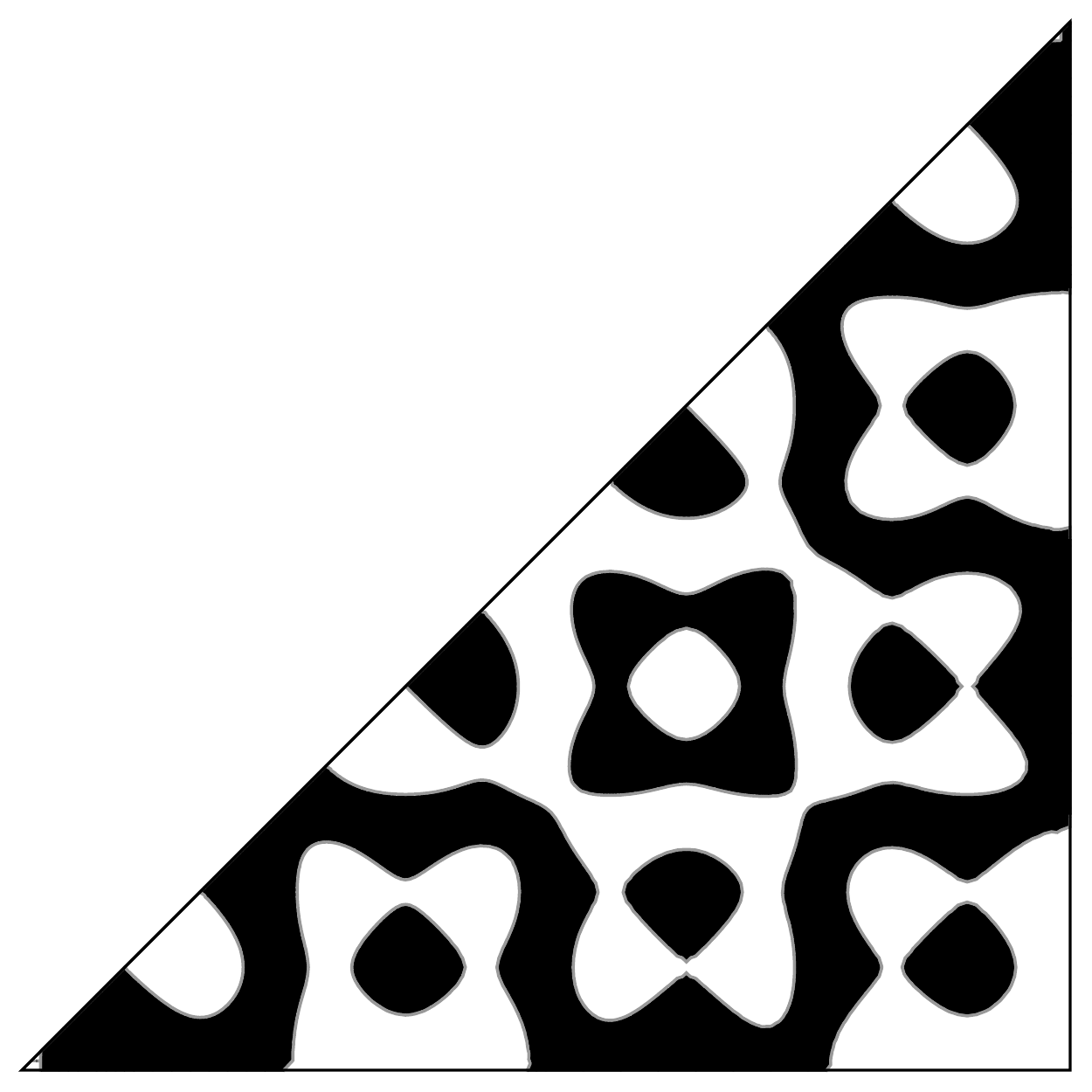}}}
\qquad \qquad
\subfloat[]{\scalebox{0.3}{\includegraphics{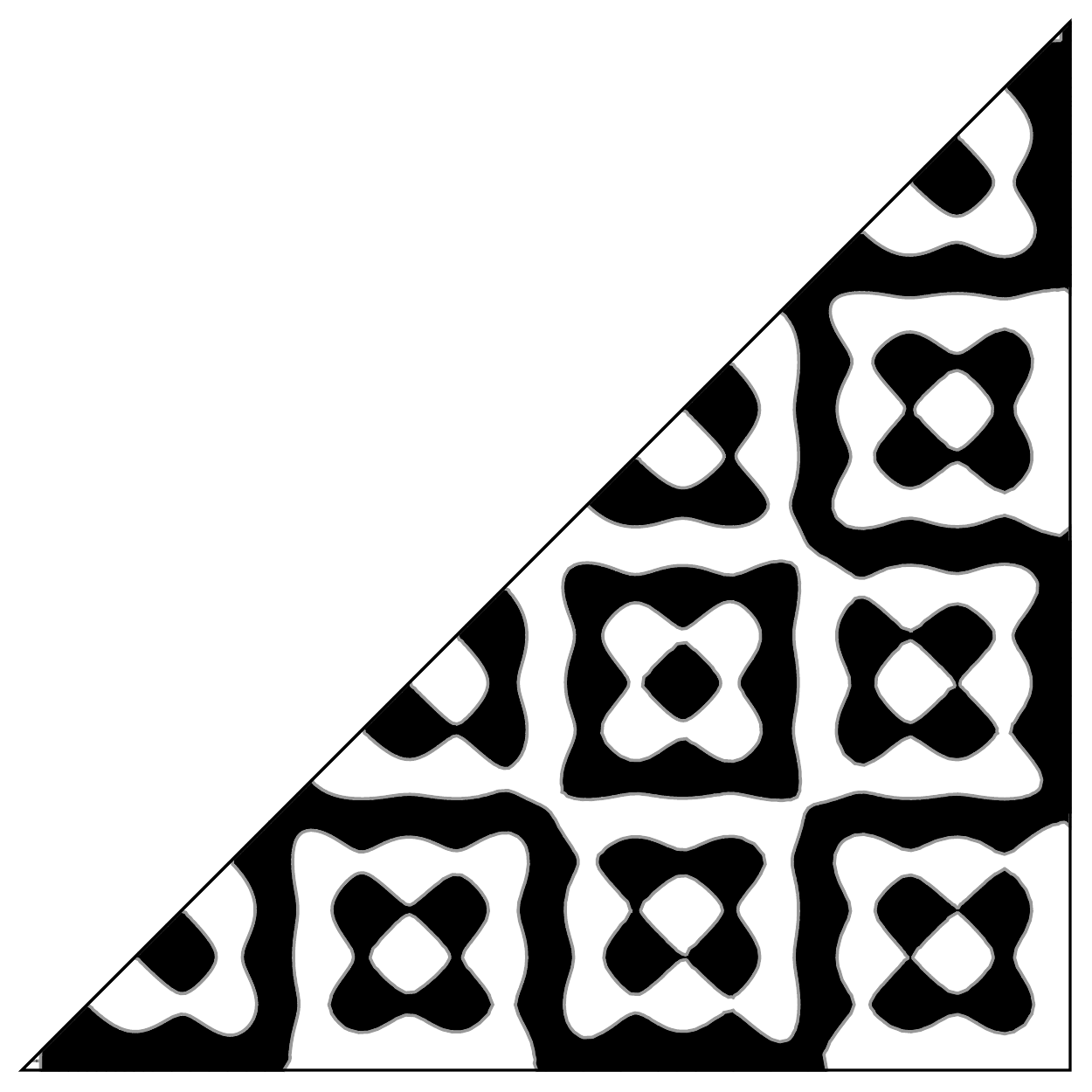}}}
\end{center}
\caption{\footnotesize The pattern of nodal domains of the right-angled isosceles triangle for (a) $\psi_{\,7,4}$, (b)  $\psi_{\,15,4}$ and (c)  $\psi_{\,23,4}$. All three eigenfunctions belong to the same equivalence class $\big[{\cal C}_{2n}\big]$ and the similarity of the nodal pattern is evident as the wavefunction evolves from one state to another within members of the same class. }
\label{examples}
\end{figure}
The right-angled isosceles triangle having the length of each equal sides as $\pi$ may be chosen as ${\cal D} = \big \{ (x, y) \in [0, \pi]^2: y \le x \big\}$. The corresponding eigenfunctions are
\begin{equation}
\psi_{m, n} (x, y) = \sin (mx)\, \sin (ny) - \sin (nx)\, \sin (my)
\label{iso}
\end{equation}
with the spectrum being determined by $E = m^2 + n^2$. Since the wavefunction is a sum of two terms, the corresponding equivalence classes are defined by $[{\cal C}_{2n}]$, where ${\cal C}_{2n} = m \mod 2n $. The total number of nodal domains $\nu_{m, n}$ can be expressed in terms of the number of nodal loops (nodal curves which neither touch the boundary nor intersect themselves or any other nodal line), $I_{m, n}$, and the total number of intersections of the nodal set with the boundary $\partial D, \: \eta_{m, n}$, as $\nu_{m, n} = 1 +\frac{1}{2}\eta_{m, n} + I_{m, n}$, where $\eta_{m, n} = m + n - 3$. The nodal domains form a tiling structure when $\displaystyle{\gcd (m, n) \ne 0 \mbox { or } m + n \equiv 0 \mod 2.}$ An empirical recursive formula for $\nu_{m, n}$ has been proposed by \cite{aronovitch}. However, observations of evaluated counts of nodal domains indicate the existence of the simpler recurrence relations
\begin{equation}
\nu_{m+2n, n} - \nu_{m, n} = \frac{n^2 + n}{2} \hspace{1cm}\mbox{ and } \hspace{1cm} I_{m+2n, n} - I_{m, n} = \frac{n^2 - n}{2}.
\label{isosceles}
\end{equation}
Let $\zeta_1 = n \mod {\cal C}$$_{2n}$ and $\zeta_2 = n \mod 2{\cal C}_{2n}$. The functional solutions to Eq. \eqref{isosceles} are given by $\nu_{m, n} = \frac{1}{4}m (n+1) + \alpha$, where $\alpha$ is a parameter that depends on ${\cal C}_{2n}$ and $n$.  Comparison of this solution with the data predicted by the recursive formula of \cite{aronovitch} renders the exact expression for the number of domains, when ${\cal C}_{2n}$ is even, as
\begin{equation}
\nu_{m, n} = \frac{m(n+1) + n - 2}{4} + \bigg[-\frac{n^2}{4} + \bigg(\frac{{\cal C}_{2n}}{2}\bigg)n - \bigg\{ \frac{{\cal C}_{2n}^2 - {\cal C}_{2n} - 1}{2} \pm \frac{1}{4} (\zeta_2 -1) \bigg\} \bigg],
\label{isosceles_nu}
\end{equation}
with the $+$ sign being applicable when ${\cal C}_{2n} < \zeta_2$ and the $-$ sign otherwise. When ${\cal C}_{2n}$ is odd, this equation is to be modified to 
\begin{equation}
\nu_{m, n} = \frac{m(n+1) + n - 2}{4} + \bigg[-\frac{n^2}{4} + \bigg(\frac{{\cal C}_{2n}}{2}\bigg)n - \bigg\{ \frac{2{\cal C}_{2n}^2 - {\cal C}_{2n} - 2}{4} + \gamma \bigg\} \bigg],
\label{isosceles_odd}
\end{equation}
where $\gamma$ is a term responsible for small fluctuations. When $\zeta_1 = 1, \: \gamma = 0$ and for $\zeta_1 = {\cal C}_{2n}- 1, \:\, \gamma$ exactly reduces to $\frac{1}{2}({\cal C}_{2n}-1)$. Although $\gamma$ increases with $n$, it remains constant within $[{\cal C}_{2n}]$ for a particular value of $n$ and hence $\displaystyle{\lim_{k \to \infty} \frac{\gamma}{\nu_{m + kn, n}} = 0}$. The explicit functional form of $\gamma$, however, remains undefined. \newline
\setlength{\tabcolsep}{16.75pt}
\renewcommand{\arraystretch}{2.5}
{\small
\begin{center}
\begin{longtable}{c c c | c c c c }
\hline
$m$ & $n$ & ${\cal C}_{2n} = m\mod 2n$ & $\nu_{m,n}$ &  $\Delta_{2n} \: \nu\,(m, n)$ & $I_{m,n} $ & $\Delta_{2n} \: I\,(m, n)$\endhead \hline
38 & 13 & 12 & 103 &   --  & 78 &   -- \\
64 & 13 & 12 & 194 & 91 & 156 & 78\\
90 & 13 & 12 & 285 & 91 & 234 & 78\\
116 & 13 & 12 & 376 & 91 & 312 & 78\\
142 & 13 & 12 & 467 & 91 & 390 & 78\\
\hline
\caption{An illustration of the constancy of the first difference of $\nu_{m, n}$ for the wavefunctions of the right-angled isosceles triangle belonging to the same class, defined by $m\mod2n$, as predicted by Eq. \eqref{isosceles}. The exact agreement of $\nu_{m, n}$ with the theoretical estimate of Eq. \eqref{isosceles_nu} is to be noted. }
\end{longtable}
\end{center}
}
\vskip -10cm
\subsection{The equilateral triangle}
Let ${\cal D} \subset \mathbb{R}^2$ be the equilateral triangle of area $\displaystyle {\cal{A}} = \frac{\sqrt{3}\pi^2}{4}$ represented as 
\begin{eqnarray}
{\cal D} = \bigg\{(x,y) \in \bigg[0,\frac{\pi}{2}\bigg] \times \bigg[0,\frac{\sqrt{3}\pi}{2}\bigg] : y\le\sqrt{3}x\bigg\} \cup \bigg\{(x,y) \in \bigg[\frac{\pi}{2},\pi \bigg] \times \bigg[0,\frac{\sqrt{3}\pi}{2}\bigg] : y\le\sqrt{3}(\pi-x)\bigg\}. 
\end{eqnarray}
The eigenfunctions of the Laplace--Beltrami operator for the general equilateral triangle billiard, with $L$ being the length of each side, form a complete orthogonal basis and are given
by \cite{McCartin} as
\begin{eqnarray}
\label{eq: wf}
\psi_{m,n}^{c,s} (x,y) &=& (\cos,\sin)\bigg[ (2m-n)\frac{2\pi}{3L}x \bigg] \sin{\bigg(n\frac{2\pi}{\sqrt{3}L}y\bigg)} - (\cos,\sin)\bigg[(2n-m)\frac{2\pi}{3L}x\bigg]\sin{\bigg(m\frac{2\pi}{\sqrt{3}L}y\bigg)} 
\nonumber \\&+&  (\cos,\sin)\bigg[-(m+n)\frac{2\pi}{3L}x\bigg]\sin{\bigg[(m-n)\frac{2\pi}{\sqrt{3}L}y\bigg]}, 
\end{eqnarray}
where $m$ and $n$ are restricted to values such that $m\geq2n$ ($m,n>0$) and the spectrum of eigenvalues is defined as $E_{m,n} = m^2+n^2-mn$. Tiling of the domains is observed when $\gcd (m, n) \ne 1$.
\begin{figure} [H]\scriptsize
\begin{center}
\subfloat[]{\scalebox{0.3}{\includegraphics{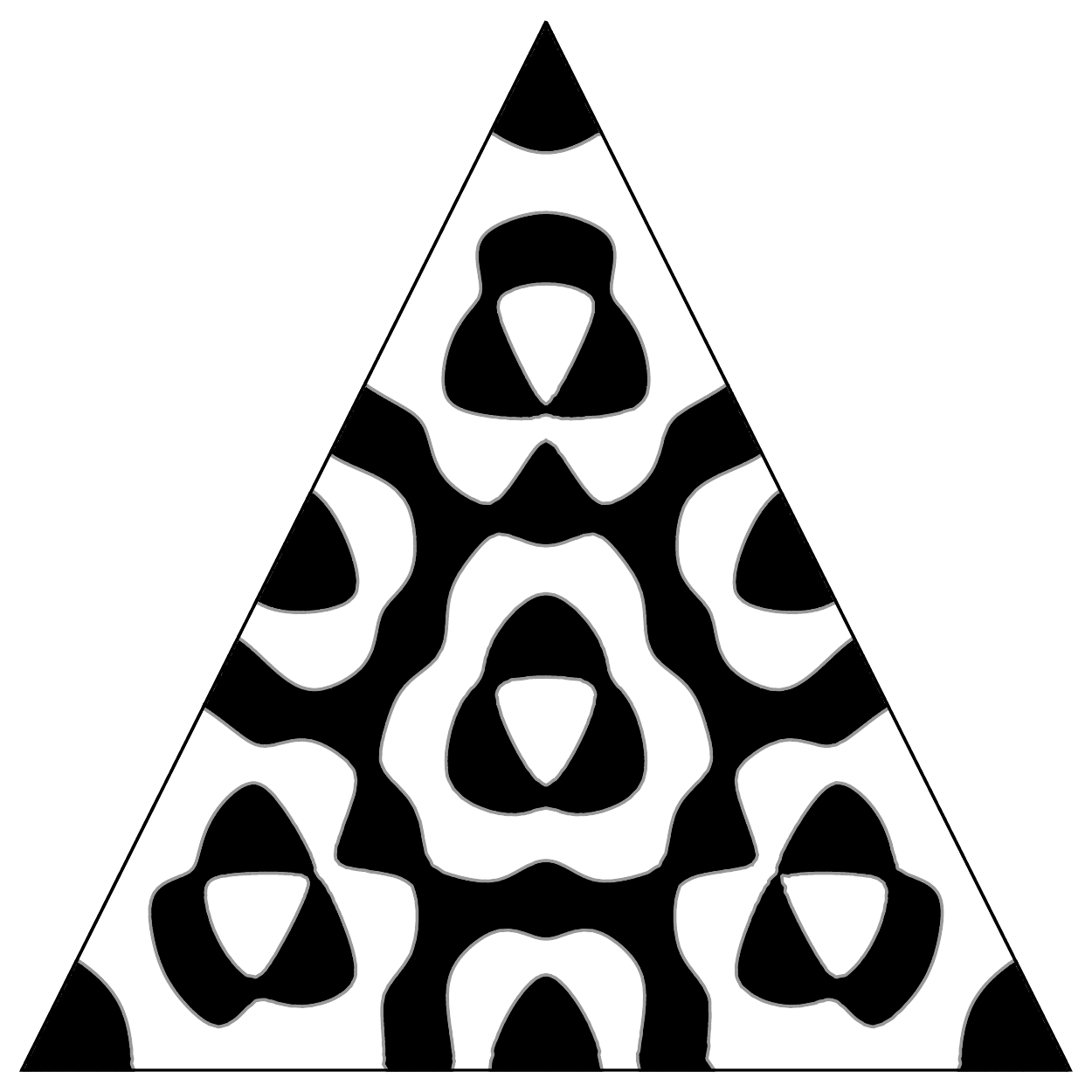}}}
\qquad \qquad
\subfloat[]{\scalebox{0.3}{\includegraphics{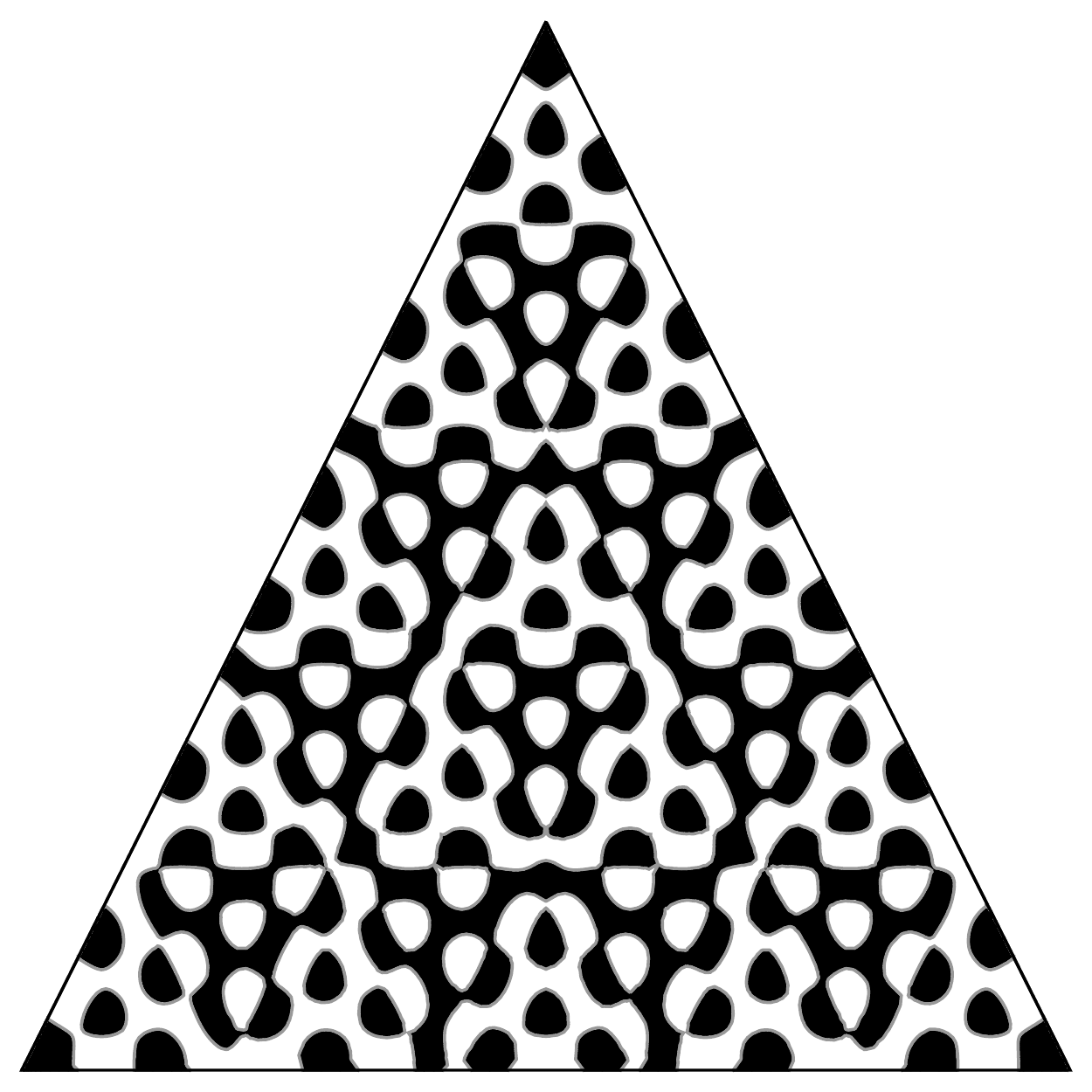}}}
\qquad \qquad
\subfloat[]{\scalebox{0.3}{\includegraphics{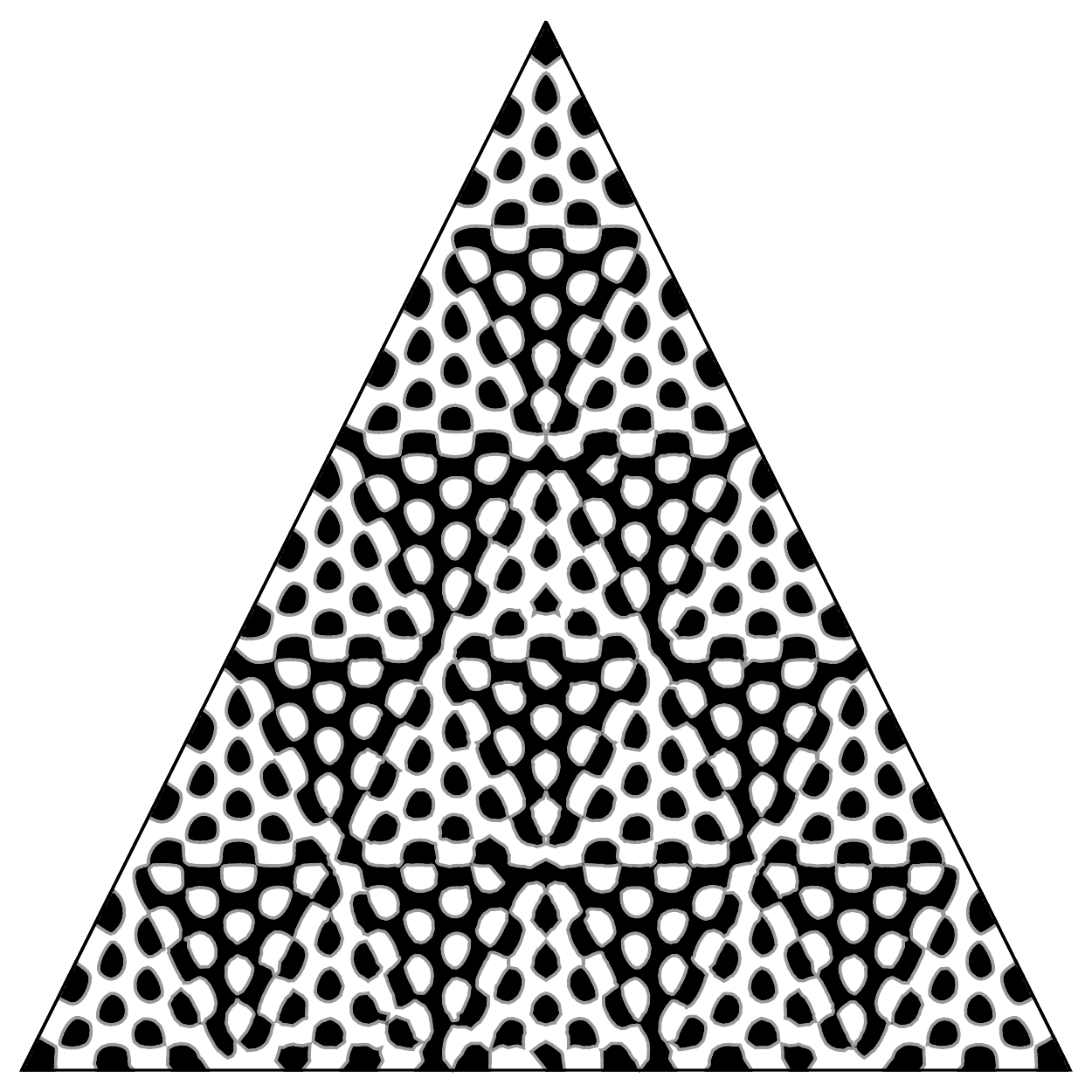}}}
\end{center}
\caption{\footnotesize The evolution of the pattern of nodal domains of the equilateral triangle from (a) $\psi_{\,16,5}$ to (b)  $\psi_{\,31,5}$ and finally to (c)  $\psi_{\,46,5}$. The geometrical similarity in the symmetry of the nodal patterns for the eigenfunctions belonging to the same equivalence class is evident.}
\end{figure}
Each non-tiling wavefunction $\psi_{m,n}$ of the symmetric mode, for a fixed value of $n$, may be categorised into one of at most $3(n-1)$ equivalence classes depending on the value of the residue ${\cal C}_{3n} = m\mod3n$. The sequence of nodal domain counts for each class, when analysed with regard to the second difference of $\nu_{m,n}$, shows the existence of the recurrence relations
\begin{equation}
\label {eq: recurrence}
\nu_{m+6n,n} - 2 \nu_{m+3n,n} +\nu_{m,n} = 3n^2  \hspace{1cm}\mbox{ and } \hspace{1cm} I_{m+6n,n} - 2 I_{m+3n,n} +I_{m,n} = 3n^2.
\end{equation}
This equation can be analytically solved to yield 
\begin{equation}
\label{eq: diff_sol}
\nu_{m,n} = \frac{3}{2} \bigg(\frac{m^2}{9} - \frac{mn}{3}\bigg) + \frac{m\alpha}{3n} + \beta,
\end{equation}
where $\alpha$ and $\beta$ are two parameters dependent on ${\cal C}_{3n}$ and $n$. Comparison of \eqref{eq: diff_sol} with the tables of evaluated domain counts clearly indicates $\displaystyle \alpha = \frac {(3n -
n^2)}{2}$. This, when coupled with further observations about the nature of the parameter $\beta$, helps to effectively reduce \eqref{eq: diff_sol} into two cases:
\begin{eqnarray}
\label{eq: approx}
\nu_{m,n} &=& \frac{m^2}{6} - \frac{(4n-3)m}{6} + n^2 - \frac{{\cal C}_{3n}n - \lambda_1({\cal C}_{3n}, n)}{3} \hspace{1.5cm} \mbox{if} \hspace{0.3cm} 0 < {\cal C}_{3n} < n,
\nonumber\\&=& \frac{m^2}{6} - \frac{(4n-3)m}{6} + n^2 - \frac{2({\cal C}_{3n} - n)n - \lambda_2({\cal C}_{3n}, n)}{3} \hspace{0.4cm} \mbox{if} \hspace{0.3cm} n < {\cal C}_{3n} < 3n.
\label{eq1}
\end{eqnarray}
Although the general forms of $\lambda_1$ and $\lambda_2$, which contribute to small variations in the nodal domain count, are unknown, it has been observed that 
\begin{equation}
\lambda_1({\cal C}_{3n}, {\cal C}_{3n}+1) = {\cal C}_{3n}^2 + 3 \qquad \mbox{ and } \qquad \lambda_2({\cal C}_{3n}, 2{\cal C}_{3n}+1) = \lambda_2({\cal C}_{3n}, 2{\cal C}_{3n}+2) = {\cal C}_{3n}({\cal C}_{3n}+3). 
\label{eq2}
\end{equation}The functional relations satisfied by $\lambda_1$ and $\lambda_2$ are further expounded in detail in \cite{sj}. 
\setlength{\tabcolsep}{8.5pt}
\renewcommand{\arraystretch}{2.5}
{\small
\begin{center}
\begin{longtable}{c c c | c c c c c }
\hline
$m$ & $n$ & ${\cal C}_{3n} = m\mod 3n$ & $\nu_{m,n}$ &  $\Delta_{3n} \nu\,(m,n)$ & $I_{m,n} $ & $\Delta_{3n} I\,(m,n)$ & $\Delta^2_{3n} \nu\,(m,n) = \Delta^2_{3n} I\,(m,n)$\endhead \hline
24 & 7 & 3 & 44 &   --  & 21 &   --  &   --  \\
45 & 7 & 3 & 198 & 154 &  154 & 133 &   -- \\
66 & 7 & 3 & 499 & 301 & 434 & 280 & 147\\
87 & 7 & 3 & 947 & 448 & 861 & 427 & 147\\
108 & 7 & 3 & 1542 & 595 & 1435 & 574 & 147\\
\hline
\caption{An example showing the values of the second difference of $\nu_{m, n}$ for the wavefunctions on the equilateral triangle corresponding to the same class, defined by $m\mod3n$, as predicted by Eq. \eqref{eq: recurrence}. The exact agreement of $\nu_{m, n}$ with the theoretical value calculated by combining equations \eqref{eq1} and \eqref{eq2} is to be noted. }
\end{longtable}
\end{center}
}

\subsubsection{The $30^\circ -60^\circ -90^\circ$ hemiequilateral triangle}
\noindent
The (30, 60, 90) scalene triangle may be regarded as a special extension of the equilateral billiard as its modes are a subset of those of the latter and precisely correspond to the states of the equilateral triangle which are antisymmetric about the altitude. These modes can be obtained by choosing the sine functions for the first components of each term in the wavefunction \eqref{eq: wf}. However, the appropriate condition for tiling in this case is noted to be $ m + n \equiv 0\, (\mbox{mod } 3) $ or $ \gcd (m, n) > 1$.
\begin{figure} [H]\scriptsize
\begin{center}
\subfloat[]{\scalebox{0.35}{\includegraphics{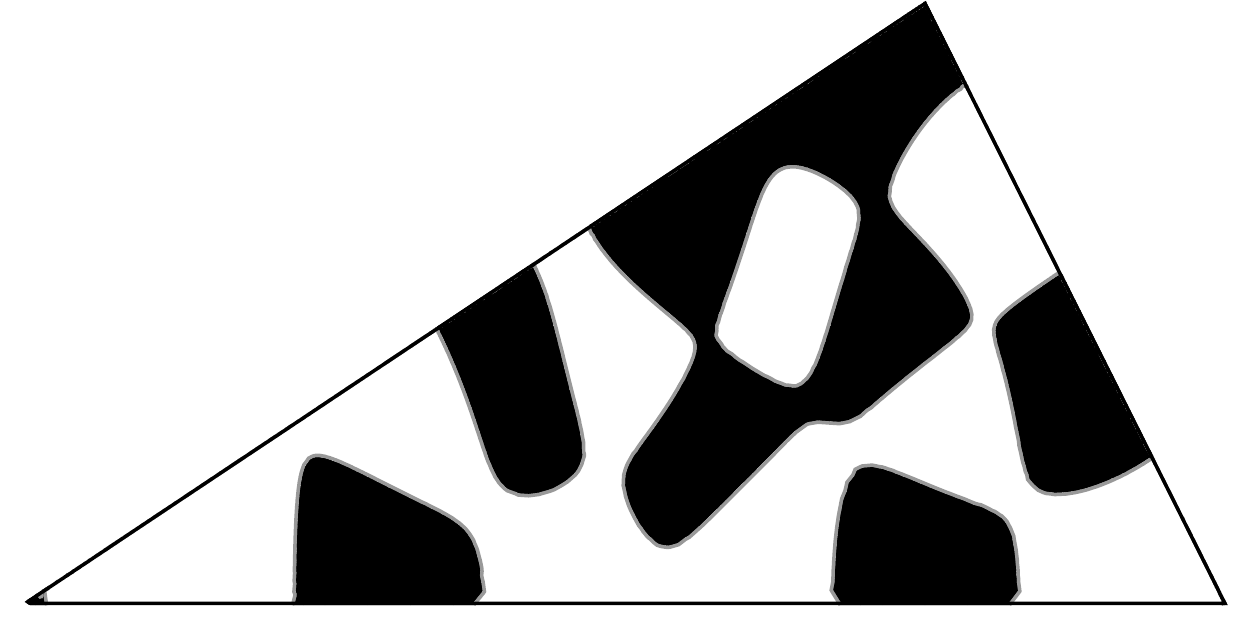}}} \quad
\subfloat[]{\scalebox{0.35}{\includegraphics{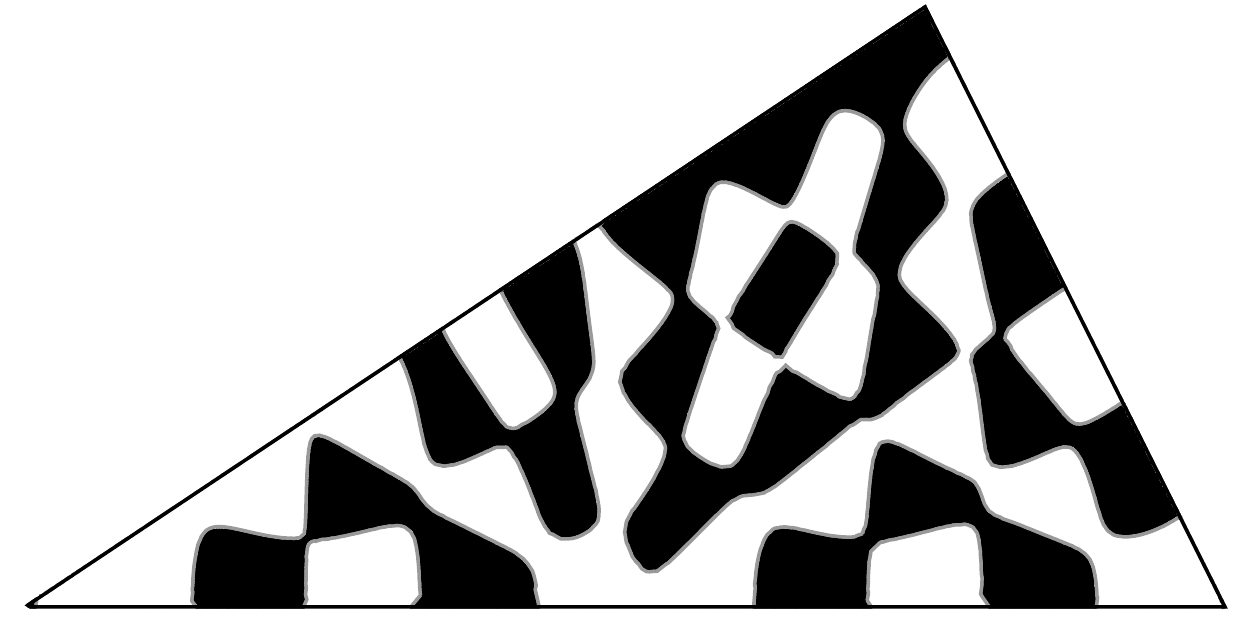}}} \quad
\subfloat[]{\scalebox{0.35}{\includegraphics{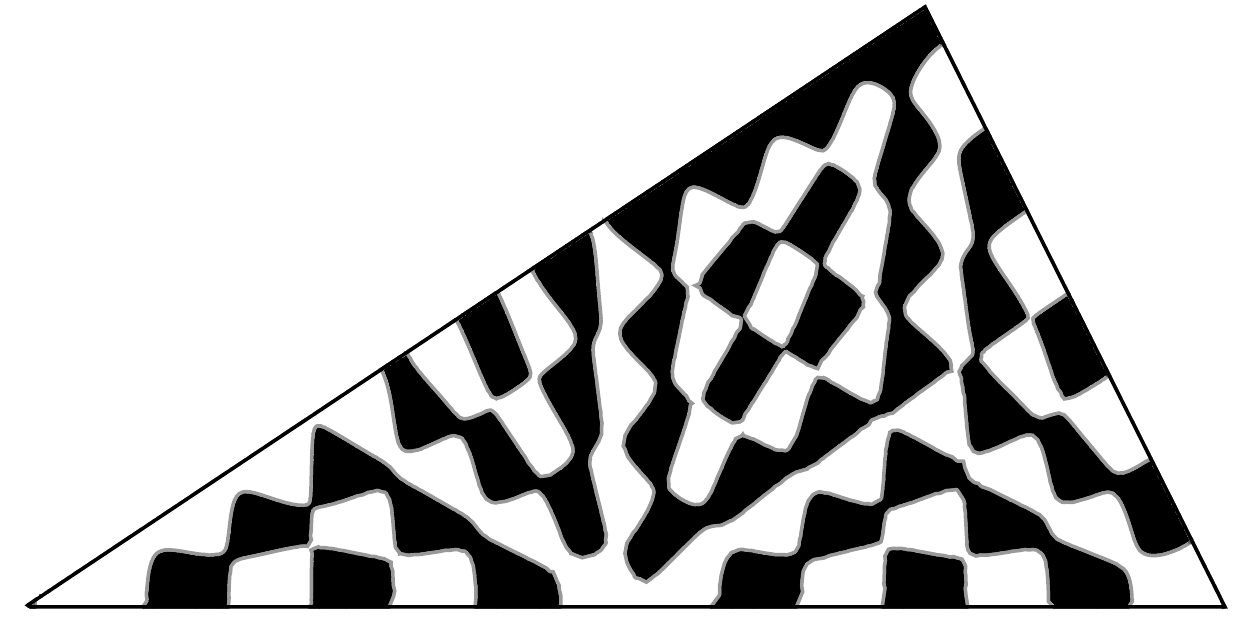}}}
\end{center}
\caption{\footnotesize The nodal domains of the (30, 60, 90) triangle for (a) $\psi_{\,11,2}$, (b)  $\psi_{\,17,2}$ and (c)  $\psi_{\,23,2}$. The self-similarity between the nodal patterns as the wavefunction evolves across different states in the same equivalence class is consistent with that observed previously for the equilateral and right-isosceles triangles.}
\end{figure}
Observations stemming from the numerical counting of the nodal domains suggest that the appropriate difference equation obeyed by the (30, 60, 90) triangular billiard is
\begin{equation}
\label {eq: scalene}
\Delta_{3n}^2 \: \nu\,(m, n) = \nu_{m+6n,n} - 2 \nu_{m+3n,n} +\nu_{m,n} = 0.
\end{equation}

\setlength{\tabcolsep}{13.5pt}
\renewcommand{\arraystretch}{2.5}
{\small
\begin{center}
\begin{longtable}{c c c | c c || c c c | c c }
\hline
$m$ & $n$ & ${\cal C}_{3n}$ & $\nu_{m,n}$ &  $\Delta_{3n} \: \nu\,(m, n)$ & $m$ & $n$ & ${\cal C}_{3n}$ & $\nu_{m,n}$ &  $\Delta_{3n} \: \nu\,(m, n)$ \endhead \hline
5 & 2 & 5 & 2 &  --  & 7 & 3 & 7 & 3 &   --  \\
11 & 2 & 5 & 7 & 5 & 16 & 3 & 7 & 13 & 10 \\
17 & 2 & 5 & 12 & 5 & 25 & 3 & 7 & 23 & 10 \\
23& 2 & 5 & 17 & 5 & 34 & 3 & 7 & 33 & 10 \\
29& 2 & 5 & 22 & 5 & 43 & 3 & 7 & 43 & 10 \\
\hline
\caption{An illustrative sample of data of the first difference of $\nu_{m, n}$ for the wavefunctions of the (30, 60, 90) triangle belonging to the same equivalence class, defined by $m\mod3n$, which displays the existence of the difference equations for this scalene triangle as conjectured by Theorem 1. }
\end{longtable}
\end{center}
}
The constancy of the first difference implies that the number of domains scales linearly as the quantum number $m$, which presents a marked departure from the equilateral triangle for which the corresponding dependence is quadratic.  The fine interconnections of nodal lines between adjacent segments of domains not only complicate the manual counting of excited states but also render the billiard unsuitable for the application of the Hoshen-Kopelman algorithm \cite{hk} and therefore, it is difficult to surmise an exact formula for the first difference. However, our extensive analysis of a considerable number of lower states enables us to estimate that $\Delta_{3n} \: \nu (m, n) \approx n^2 + 1$ for the non-tiling situations.

\section{Conclusions}

Plane polygonal billiards with internal angles $\pi /n_k$ are integrable as the resulting shapes tessellate the plane and all the polygonal domains possessing a complete set of trigonometric eigenfunctions of the Laplacian, under either Dirichlet or Neumann boundary conditions, are documented by \cite{mc}. The only triangular solutions are the right isosceles, equilateral, and the (30-60-90) triangles. In addition, the rectangular billiard is integrable. Non-polygonal, convex shapes are integrable if the Helmholtz equation can be separated in an appropriate coordinate system. For instance, the circular and elliptical billiards are integrable. In this article, for \textsl{all} the above-mentioned systems, we have shown that the number of domains, $\nu _{m, n}$ of an eigenfunction satisfies a difference equation. Since the arguments given for the circle and the ellipse are quite general, it is expected that $\nu _{mn}$ for other separable planar billiards satisfy similar equations. As classifying patterns in non-separable shapes and counting domains has been a very difficult problem, the theorem presented here marks a considerable advance. In short, \textsl{the geometrical patterns have been algebraically represented}.

Perhaps the next study will be on the simplest non-integrable system -- the $\pi /3$-rhombus billiard \cite{jain1,jain2}. The eigenfunctions of this system cannot be represented as random waves \cite{jain3}. The results presented here and recently found for the equilateral triangle billiard \cite{sj} invite a similar study for the rhombus billiards. However, this ``non-integrable step" seems quite challenging. We hope that the present result which is, in fact, a two-dimensional generalization of Courant's work for a one-dimensional system, would suggest the nontrivial step without referring to the random-wave hypothesis. We believe that the connection with a certain class of difference equations might play a significant role in realising the next step. 

\appendix
\section{Explanation of the similarity of nodal patterns in $[{\cal C}_{2n}]$}
Consider the eigenfunction of the right-angled isosceles triangular billiard given by \eqref{iso}. We examine the origin of the similarity in the nodal domain patterns of $\psi_{m, n} (x, y)$ and $\psi_{m+2n, n} (x, y)$ that was illustrated in Figure \ref{examples}. We believe that a geometric argument for the similarity between these two states would naturally explain the similarity between all states belonging to the same equivalence class $[{\cal C}_{2n}]$ thereafter.
\begin{eqnarray}
\label{app1} \psi_{m+2n, n} (x, y) &=& \sin [(m+2n)x] \sin (ny) - \sin(nx) \sin [(m+2n)y]\\
\nonumber&=&\kappa \sin(mx)\sin(ny) - \chi\sin(nx)\sin(my) + \cos(mx)\sin(2nx)\sin(ny) - \cos(my)\sin(nx)\sin(2ny),
\end{eqnarray}
where, $\kappa = \cos (2nx) \mbox{ and } \chi = \cos (2ny)$, which implies that $\lvert \kappa \rvert$,  $\lvert \chi \rvert < 1.$ Using the notation
\begin{equation}
\tau_{m, n} (x, y) = \kappa \sin(mx)\sin(ny) - \chi\sin(nx)\sin(my),
\label{app2}
\end{equation}
which groups together the first two terms in the expansion of \eqref{app1}, we see that $\tau_{m, n} (x, y)$ represents a linear combination of the two terms of $\psi_{m, n}$. However, the weights assigned to each term of the sum is $\ne 1$ as in the eigenfunction \eqref{iso}, but rather $\kappa$ and $\chi$ respectively. The distortion of the nodal lines of the wavefunction $\theta_1\sin(mx)\sin(ny) + \theta_2\sin(nx)\sin(my)$ from those of the original wavefunction ($\theta_1 = \theta_2 = 1$) with the variation of the parameters $\theta_1$ and $\theta_2$ has  been discussed and diagrammatically presented in \cite{courant}. This perturbed equation is analogous to $\tau_{m, n} (x, y)$, except for the fact that the weights in the latter, albeit bounded, are functions of $x$ and $y$ and are therefore, position-dependent. Nevertheless, it is observed that the resultant nodal domains are not greatly distorted from the original domains of the triangular billiard and the initial and perturbed nodal lines are also comparable. This may be accounted for by the reasoning that when $\kappa = \chi$, $\tau_{m, n} (x, y) = \kappa\, \psi_{m, n} (x, y)$ and hence $\tau_{m, n} (x, y)$ differs from the original wavefunction of the system by only a multiplicative constant, therefore leaving the structure and number of nodal domains unaffected. The alteration of the shapes of the nodal domains of the wavefunction on the billiard with the values of $\kappa$ and $\chi$ are documented in the Mathematica notebook accompanying this article. From this heuristic framework, it is evident that $\tau_{m, n} (x, y)$ preserves the basic structure of the domains of $\psi_{m,n}$ and is therefore the term that is responsible for preserving the similarity of the nodal pattern as $m \rightarrow m+2n$ .   

It is insightful to examine at this point the geometric nature of the third and fourth terms in the trigonometric expansion of \eqref{app1}. We employ the notation
\begin{eqnarray}
\label{app3}\sigma_{m, n} (x, y) &=& (\psi - \tau)_{m, n} (x, y) = \big[\cos (mx)\big]\big[\sin (2nx) \sin(ny)\big] - \big[\cos (my)\big]\big[\sin (nx) \sin(2ny)\big]\\
\label{app4}&=& 2 \sin (nx) \sin (ny) \big[\cos (nx) \cos (mx) - \cos(ny)\cos(my) \big].
\end{eqnarray}
It is apparent that \eqref{app3} resembles \eqref{app2} in the sense that it is also a linear combination of the individual terms of $\psi_{2n, n}$, with the respective weights being $\kappa' = \cos (mx) $ and $\chi' = \cos (my)$. However, since one quantum number is an integer multiple of the other, $\psi_{2n, n}$, and consequently $\sigma_{m, n} (x, y)$ corresponds to a tiling state. The zeros of \eqref{app4}, corresponding to the nodal sets of each of the three terms in the product, occur along the lines $\displaystyle{\{x = \omega \pi/n\}, \{y = \omega \pi/n\}} \mbox{ and } \{x = y\}$, where $0 \le \omega \le n$, thereby forming a grid-like nodal structure. Neither $\tau_{m, n} (x, y)$ nor $\sigma_{m, n} (x, y)$ is an eigenfunction of the triangle and hence do not necessarily satisfy the Dirichlet boundary conditions. 
\begin{figure} [H]\scriptsize
\begin{center}
\subfloat[]{\scalebox{0.3}{\includegraphics{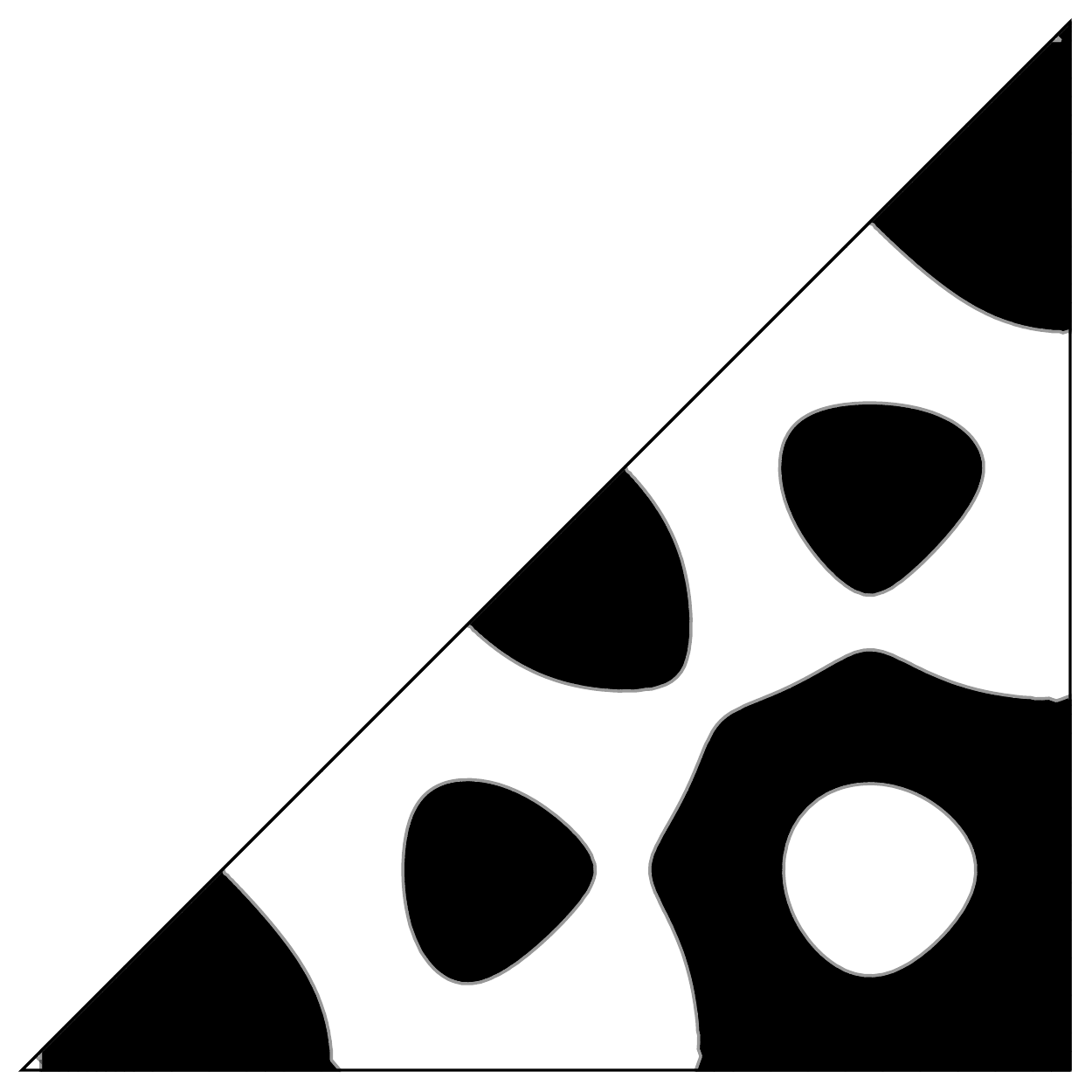}}}
\quad
\subfloat[]{\scalebox{0.3}{\includegraphics{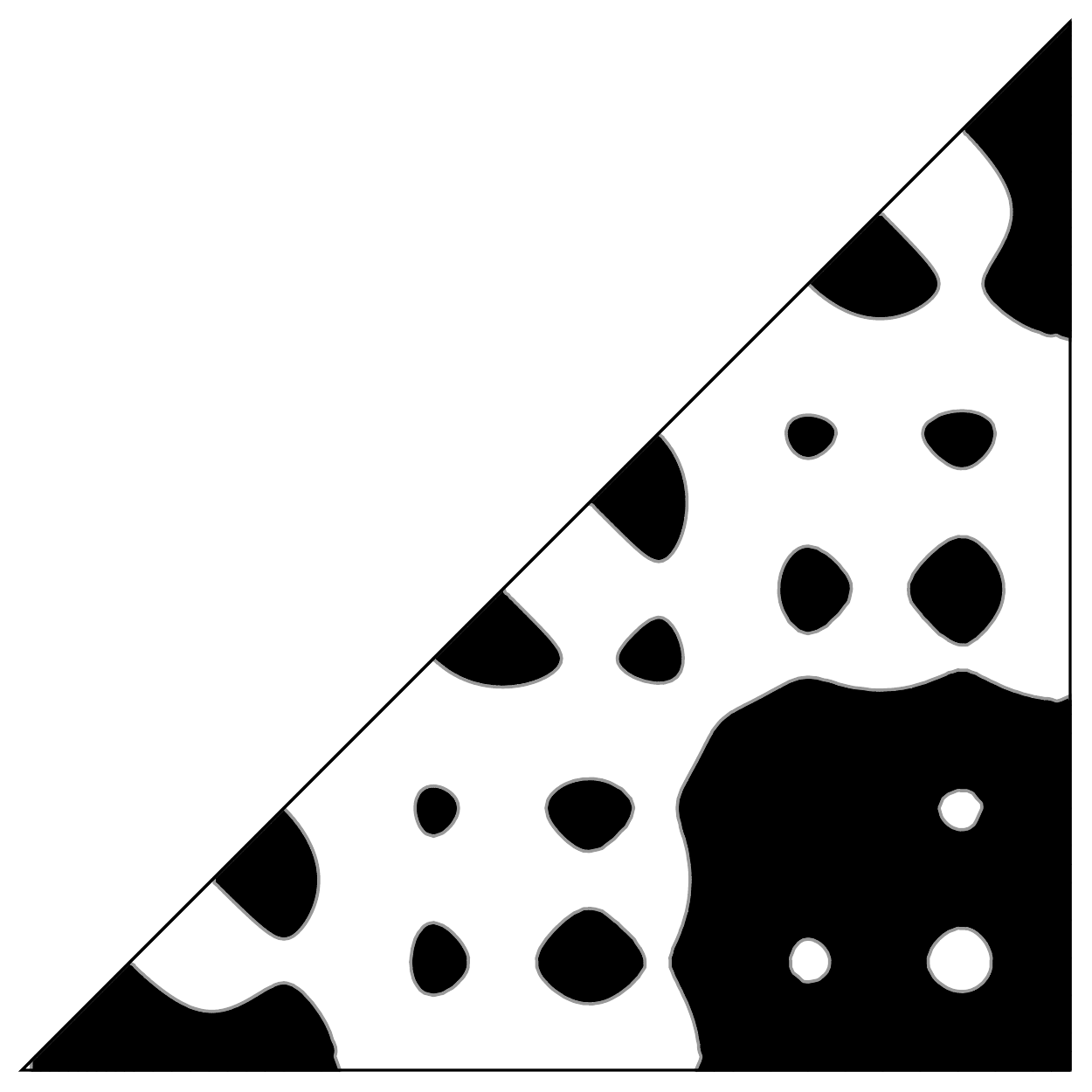}}}
\quad
\subfloat[]{\scalebox{0.3}{\includegraphics{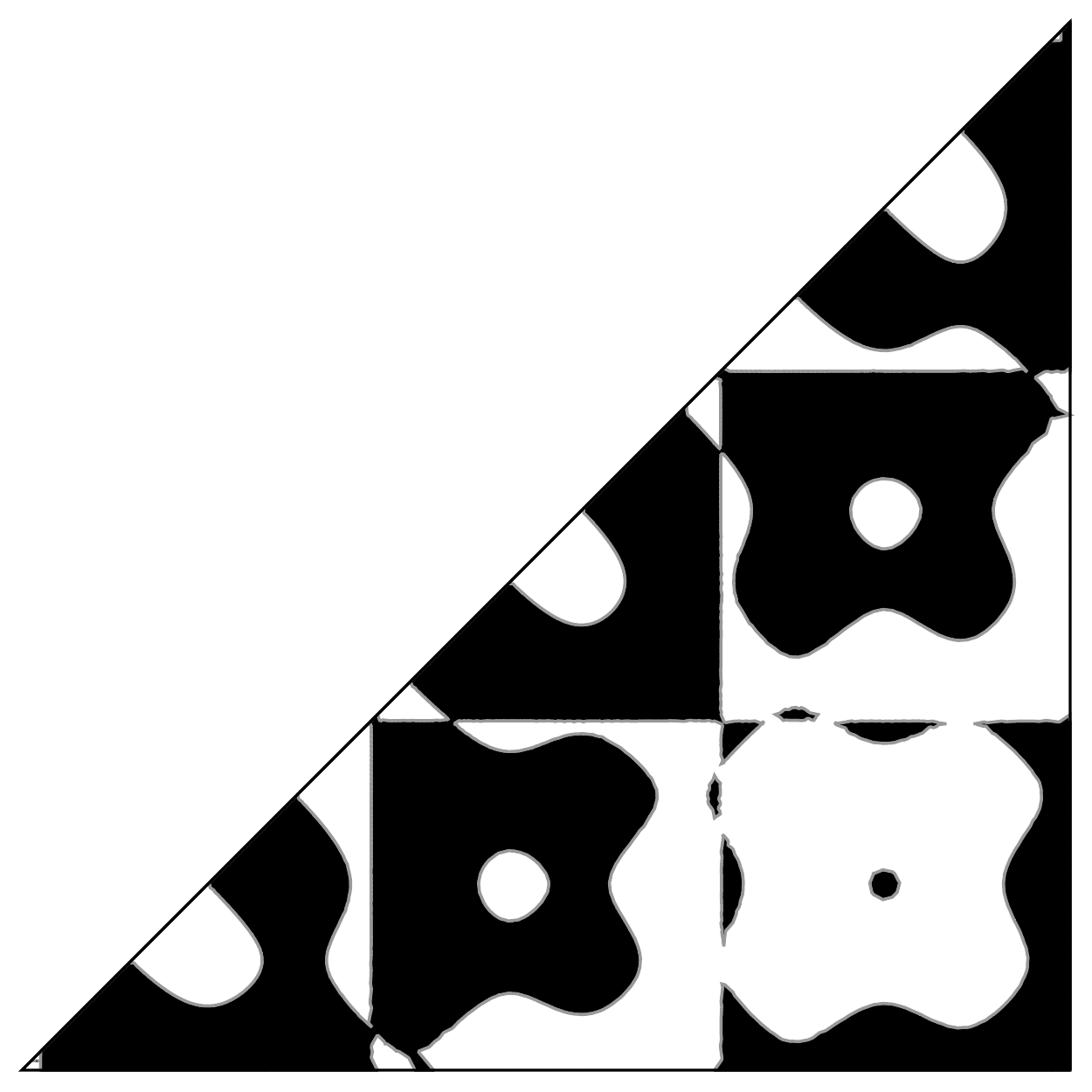}}}
\quad
\subfloat[]{\scalebox{0.3}{\includegraphics{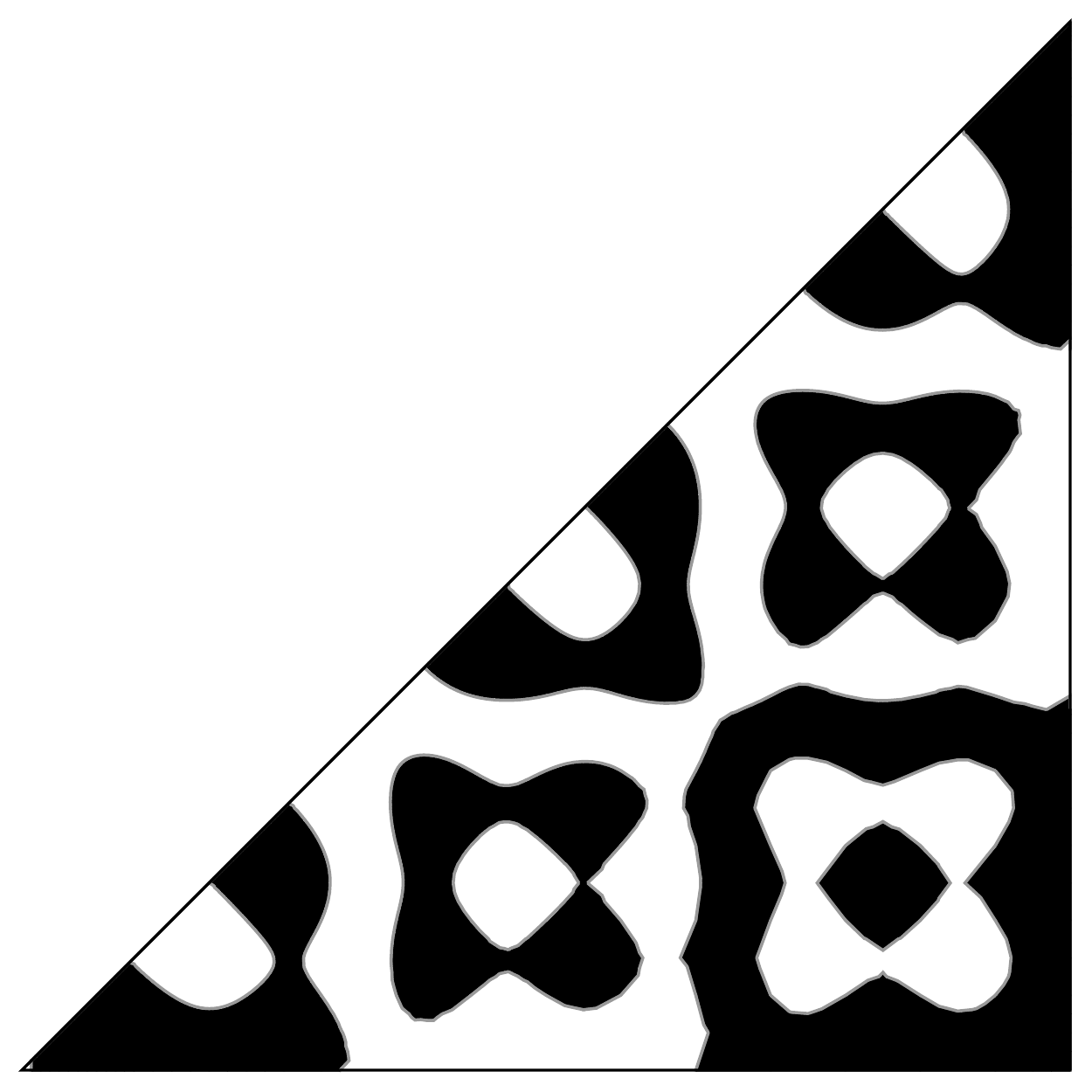}}}\\
\subfloat[]{\scalebox{0.3}{\includegraphics{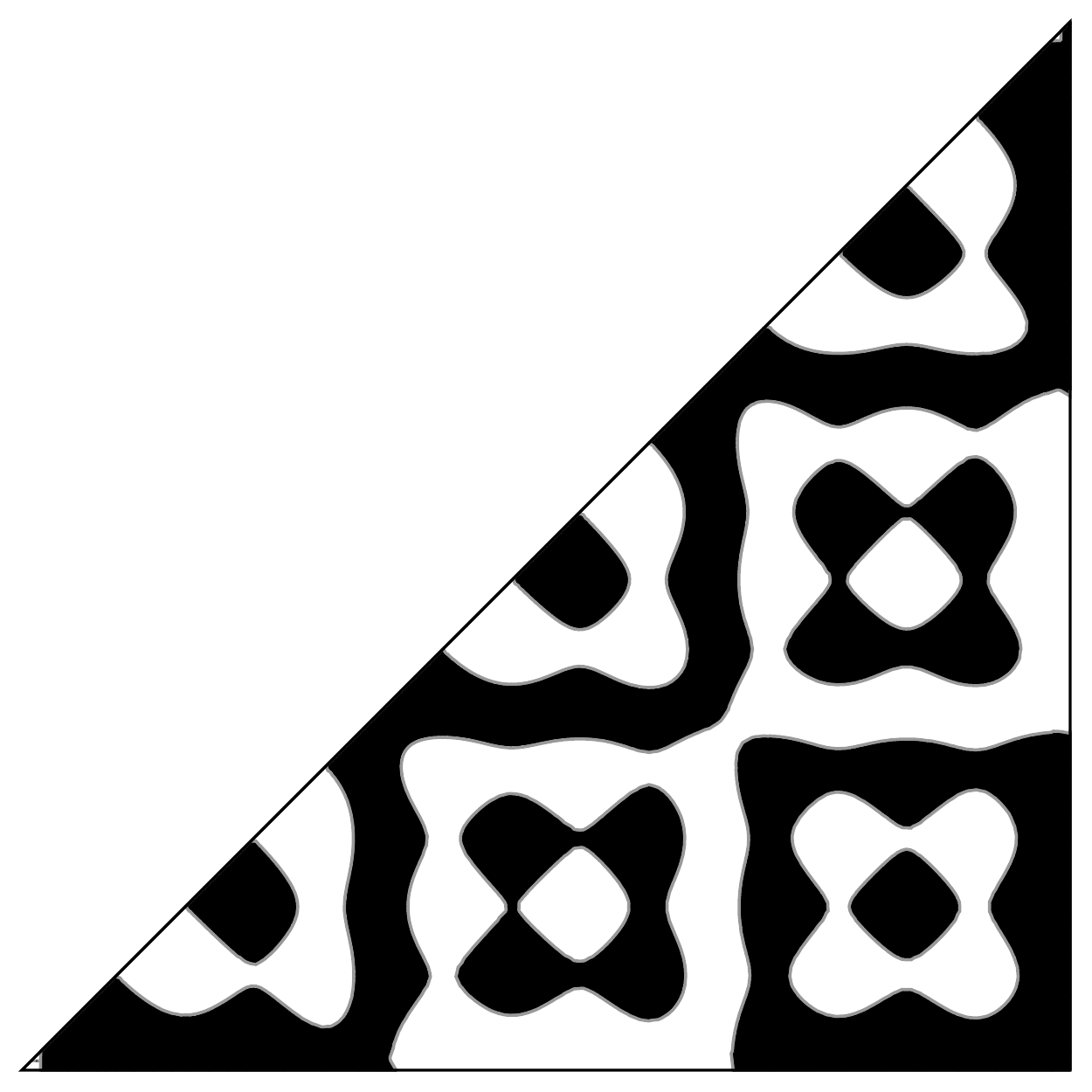}}}
\quad
\subfloat[]{\scalebox{0.3}{\includegraphics{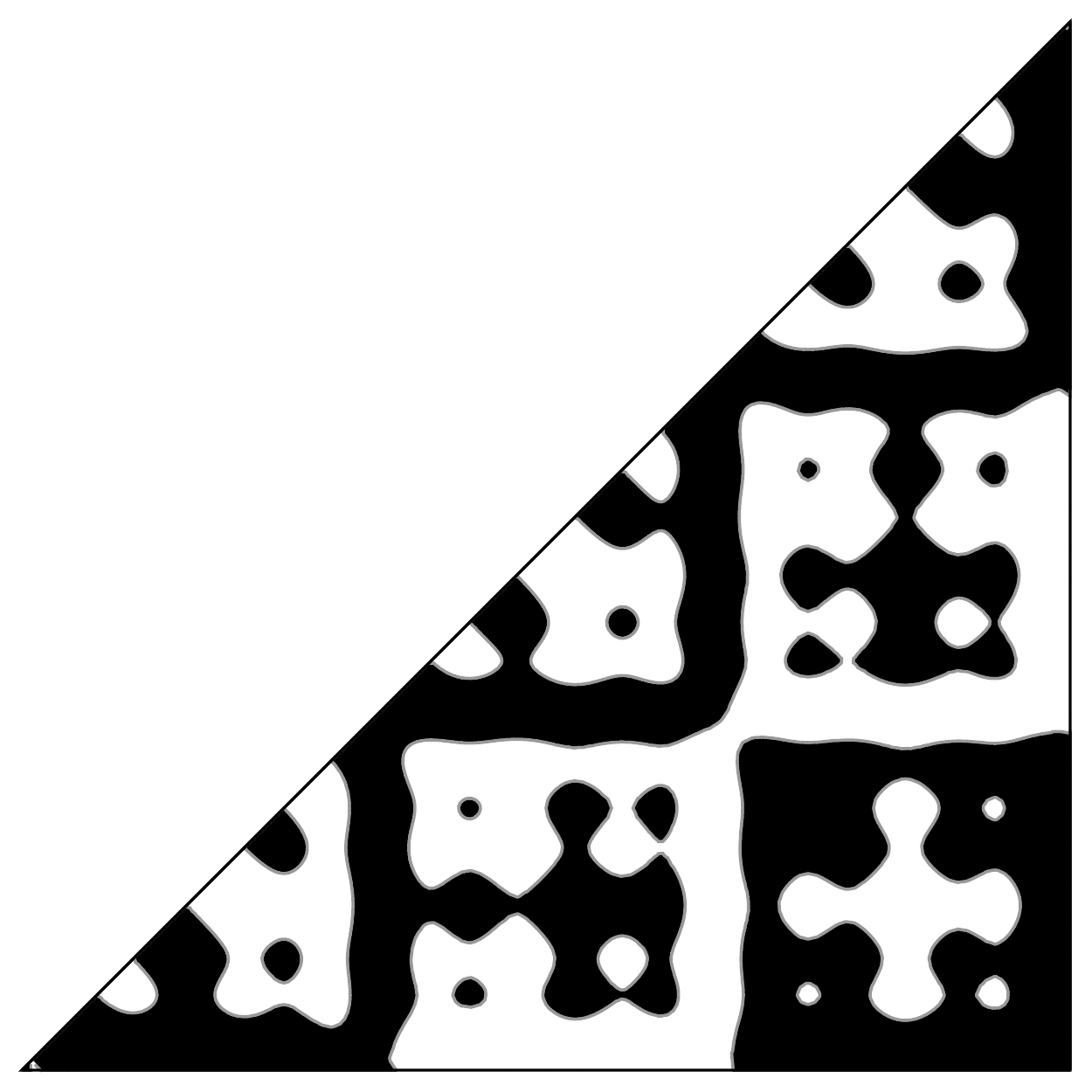}}}
\quad
\subfloat[]{\scalebox{0.3}{\includegraphics{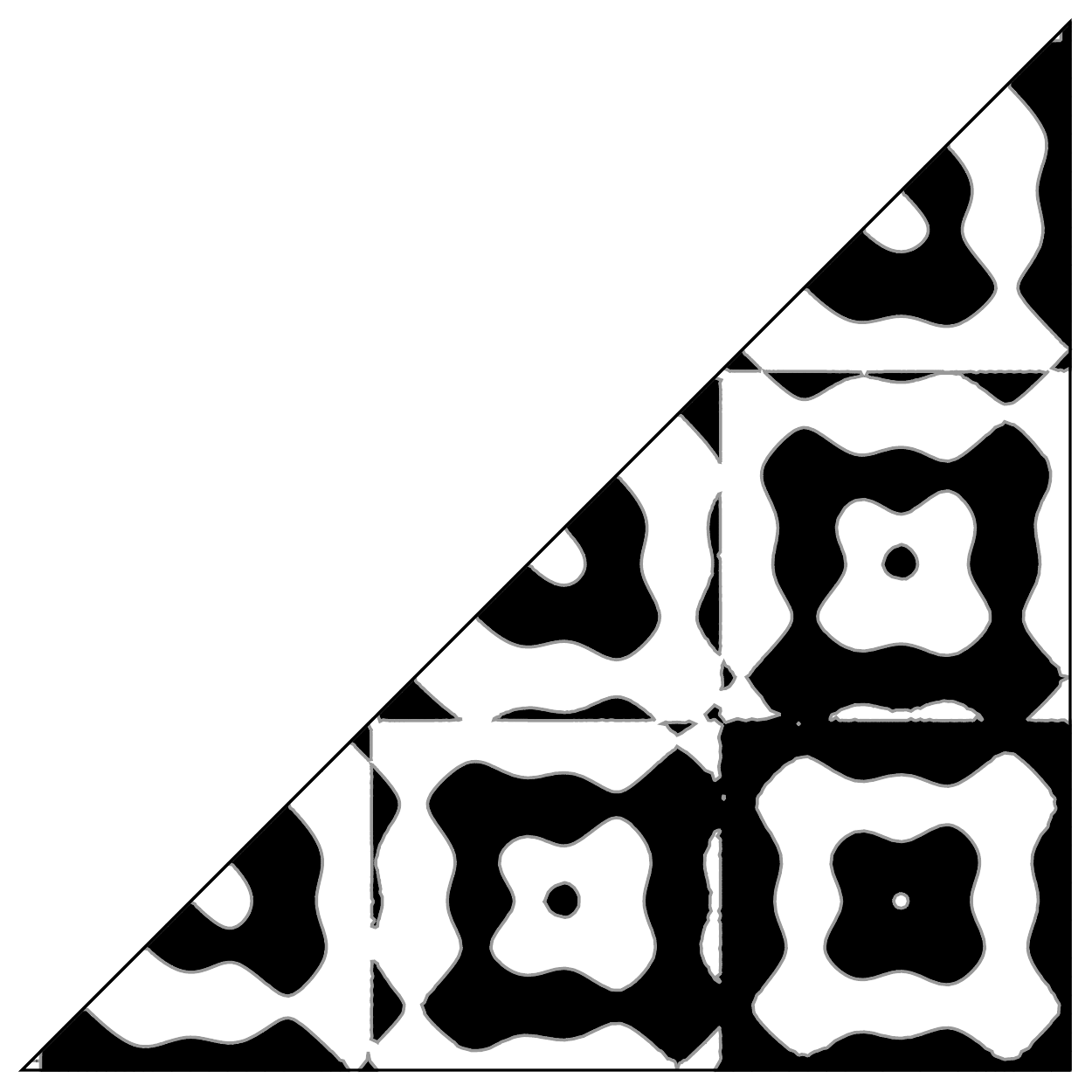}}}
\quad
\subfloat[]{\scalebox{0.3}{\includegraphics{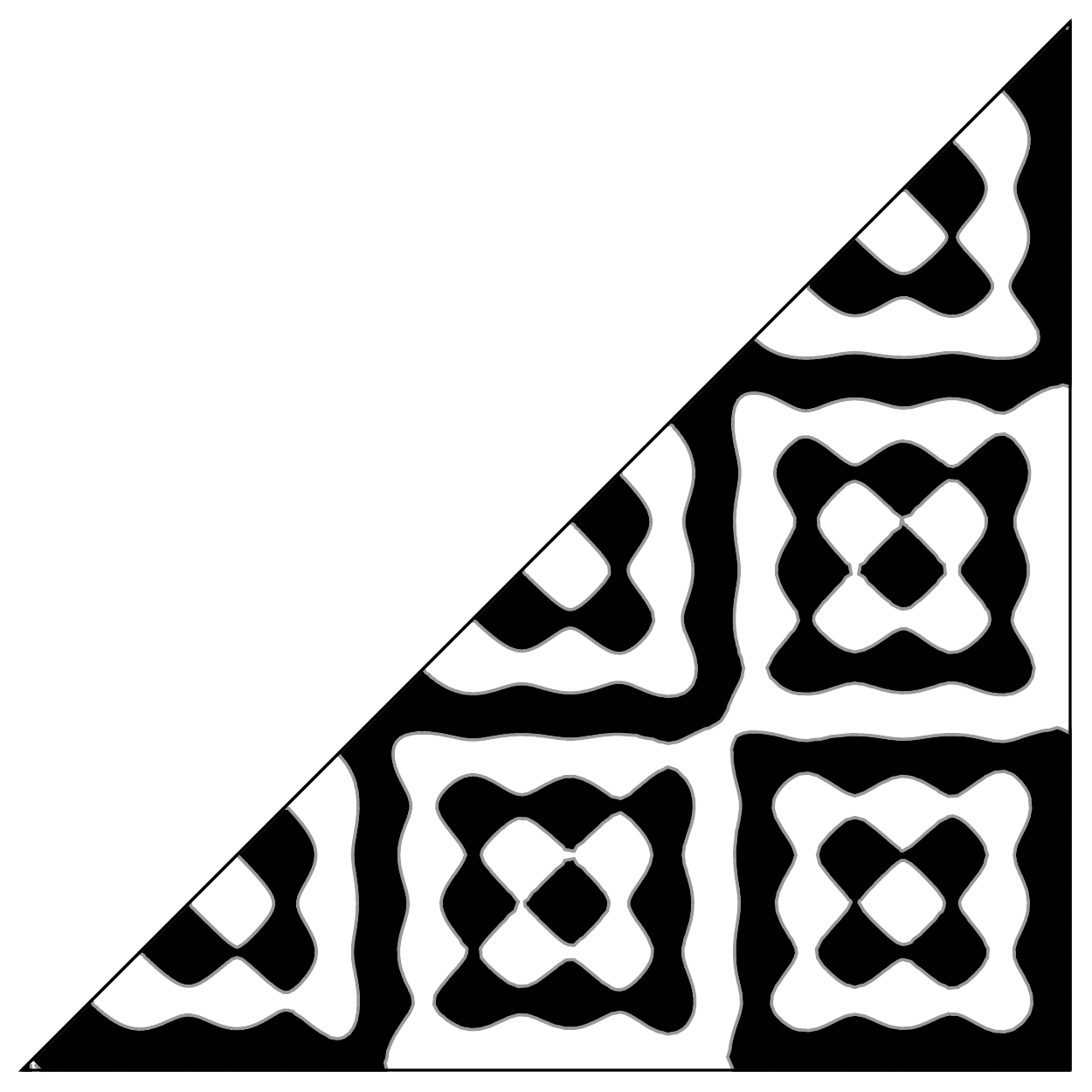}}}
\end{center}
\caption{\footnotesize The geometric similarity between the states (a) $\psi_{8, 3}$ and (d) $\psi_{14, 3} \in [2]$ can be understood by decomposing $\psi_{14, 3}$ into its (b) structure-preserving ($\tau_{8, 3}$) and (c) tiling ($\sigma_{8, 3}$) components. Similarly, decomposition into (f) $\tau_{16, 3}$ and (g) $\sigma_{16, 3}$ of (h) $\psi_{22, 3}$ elucidates its nodal pattern's similarity with (e) $\psi_{16, 3}$. }
\label{convince}
\end{figure}

As can be observed from Figure \ref{convince}, additional nodal domains arise in $\psi_{m + 2n, n}$ within regions that constituted a single, continuous domain for $\psi_{m, n}$. This is because the set 
\begin{equation*}
\Xi = \big \{ (x, y) \in [0, \pi]^2: y \le x\; \lvert\; \sgn \big(\tau_{m, n}(x, y)\big) = -\sgn \big(\sigma_{m, n} (x, y)\big)  \big\} \ne \emptyset,
\end{equation*}
and has non-zero measure. Therefore, the functions $\tau_{m, n}(x, y)$ and $\sigma_{m, n}(x, y)$ tend to cancel each other locally within the connected regions of the triangle wherein they differ in sign, leading to loss of convexity of the domain, if any. The continuity of the nodal lines for the state thus formed, $\psi_{m+2n, n}$, is ensured by Bers' theorem \cite{bers}, which states that in two dimensions, if $\Phi(\vec{z}) = 0$, then in any neighbourhood of $\vec{z}$, the nodal line is either a smooth curve or a union of $n$ smooth curves intersecting at equal angles. In fact, an implicit assumption of this theorem underlies our ansatz of the structure-preserving property of $\tau_{m, n}(x, y)$ as it enables one to conjecture that a collection of smooth curves would remain reasonably smooth under small weighted perturbations, thereby resembling the initial set of nodal lines. The region wherein the condition for the formation of a new connected domain is satisfied is:
\begin{equation}
\mathbb{U} = \mbox{int}(\Xi) \cap \big\{(x, y): \big[\sgn \big(\psi_{m,n} (x, y)\; \sigma_{m,n} (x, y)\big) \big]  \big(\lvert \tau_{m, n} (x, y) \rvert - \lvert \sigma_{m, n} (x, y) \rvert \big) > 0\big\}.
\label{condition}
\end{equation} 
Let $\Upsilon = \{ \psi_{m, n} > 0 \} \cap \mathbb{U}$, such that $\displaystyle{\Upsilon \cap \frac{1}{2} \mathbb{U} \ne \emptyset}$. A local statement analogous to Courant's nodal domain theorem \cite{donnelly} shows that
\begin{equation*}
\frac{\mbox{Area }(\Upsilon)}{\mbox{Area }(\mathbb{U})} \ge a. \lambda^{-p},
\end{equation*}
where $a$ depends only on the Riemannian metric and $p$ is a constant for a two-dimensional surface. This decay shows the scaling of the regions of the triangular billiard within which new nodal domains may form as one moves to higher excited states corresponding to higher eigenvalues $\lambda$.

The size of a nodal domain can be characterised by its in-radius $r_\lambda$ and it has been proved \cite{man} that 
\begin{equation*}
\frac{C_1}{\sqrt{\lambda}} \ge r_\lambda \ge \frac{C_2}{\lambda^{\frac{1}{4}\big(n^2 - \frac{15n}{8} + \frac{1}{4}\big)}(\log \sqrt{\lambda})^{2n - 4}},
\end{equation*}
where $C_1$ and $C_2$ are constants (not to be confused with the equivalence classes). For two dimensions, it is known that \cite{zelditch}
\begin{equation}
\frac{C_1}{\sqrt{\lambda}} \ge r_\lambda \ge \frac{C_2}{\sqrt{\lambda}}.
\label{ratio}
\end{equation}
The number of nodal domains may be crudely approximated as $\displaystyle{\nu_{m, n} \sim \frac {\mbox{Area }(\cal{D})}{\pi r_\lambda^2}}.$ From \eqref{ratio}, we have
\begin{eqnarray}
\nonumber&&\frac{\pi (m^2 +n^2)}{2C_1^2} \le \nu_{m, n} \le \frac{\pi (m^2 + n^2)}{2C_2^2} \hspace{0.5cm} \mbox{and} \hspace{0.5cm} \frac{\pi (m^2 + 5n^2 + 4mn)}{2C_1^2} \le \nu_{m +2n, n} \le \frac{\pi (m^2 + 5n^2 + 4mn)}{2C_2^2},
\\&&\therefore \frac{2\pi (mn + n^2)}{C_1^2} \le \Delta_{2n} \: \nu (m, n) = \nu_{m+2n, n} - \nu_{m, n} \le\frac{2\pi (mn + n^2)}{C_2^2}.
\label{check}
\end{eqnarray}
Thus \eqref{check} predicts that $\Delta_{2n} \: \nu (m, n) \propto n^2 + n$ and scales as $\BigO{n^2}$, which is in exact agreement with \eqref{isosceles}. We believe that exactly the same argument involving trigonometric expansion of the wavefunction may be applied to the equilateral triangle billiard to explain the geometric similarity of the nodal patterns in $[{\cal C}_{3n}]$. 

 \section*{Acknowledgements}
The authors would like to thank H. R. Krishnamurthy for an interesting discussion regarding the geometric similarity between wavefunctions belonging to the same equivalence class. Furthermore, the authors are grateful to Howard Wilson for kindly providing the MATLAB code and for his assistance when they were in the process of visualising the wavefunctions on the ellipse.

\end{document}